\documentclass[useAMS]{mn2e}
\pdfoutput=1 
\usepackage{graphicx}
\usepackage{amssymb}
\usepackage{amsmath}

\usepackage{color}

\newcommand\be{\begin{equation}}
\newcommand\ee{\end{equation}}

\newcommand{\ii}{\mathrm{i}}
\newcommand{\dd}{\mathrm{d}}
\renewcommand{\vec}[1]{\mbox{\protect\boldmath$#1$}}
\newcommand{\tens}[1]{\mathbf{#1}}

\title [Oscillations in Viscous BH Accretion Discs]{Viscous Driving of Global Oscillations in Accretion Discs Around Black Holes}
\author [R. Miranda, J.~Hor\'{a}k and D. Lai]
        {Ryan Miranda,$^1$ Ji\v{r}\'{\i}~Hor\'{a}k$^2$ and Dong Lai$^1$ \\
         $^1$Department of Astronomy, Cornell University, Ithaca, NY 14853, USA \\
         $^2$Astronomical Institute of the Academy of Sciences, Bo\v{c}n\'{\i}~II 1401/1a, 141-31 Praha~4, Czech Republic}

\begin{document}

\maketitle

\begin{abstract}
We examine the role played by viscosity in the excitation of global oscillation modes (both axisymmetric and non-axisymmetric) in accretion discs around black holes using two-dimensional hydrodynamic simulations. The turbulent viscosity is modeled by the $\alpha$-ansatz, with different equations of state. We consider both discs with transonic radial inflows across the innermost stable circular orbit, and stationary discs truncated by a reflecting wall at their inner edge, representing a magnetosphere. In transonic discs, viscosity can excite several types of global oscillation modes. These modes are either axisymmetric with frequencies close to multiples of the maximum radial epicyclic frequency $\kappa_\mathrm{max}$, non-axisymmetric with frequencies close to multiples of of the innermost stable orbit frequency $\Omega_\mathrm{ISCO}$, or hybrid modes whose frequencies are linear combinations of these two frequencies. Small values of the viscosity parameter $\alpha$ primarily produce non-axisymmetric modes, while axisymmetric modes become dominant for large $\alpha$. The excitation of these modes may be related to an instability of the sonic point, at which the radial infall speed is equal to the sound speed of the gas. In discs with a reflective inner boundary, we explore the effect of viscosity on trapped $p$-modes which are intrinsically overstable due to the corotation resonance effect. The effect of viscosity is either to reduce the growth rates of these modes, or to completely suppress them and excite a new class of higher frequency modes. The latter requires that the dynamic viscosity scales positively with the disc surface density, indicating that it is a result of the classic viscous overstability effect.
\end{abstract}


\section{Introduction}
High-frequency quasi-periodic oscillations (HFQPOs) in X-ray flux have been observed in a number of black hole X-ray binaries since the 1990s. These low-amplitude ($\sim 1\%$) variabilities occur in the so-called intermediate state or ``steep power law'' state, and have frequencies of $40 - 450$ Hz, comparable to the orbital frequency the innermost stable circular orbit (ISCO) of a $\sim 10$ $\mathrm{M}_\odot$ black hole. This indicates that the physical mechanism behind them is closely related to the dynamics of the innermost regions of black hole accretion discs (for reviews, see Remillard \& McClintock 2006 and Belloni et al. 2012). Several classes of models for HFQPOs have been proposed (see Lai \& Tsang 2009, Lai et al. 2013 for a review), including the orbital motion of hot spots in the disc (Stella et al. 1999; Schnittman \& Bertschinger 2004; Wellons et al. 2014), nonlinear resonances (Abramowicz \& Klu\'{z}niak 2001; Abramowicz et al. 2007), and oscillations of finite accretion tori (Rezzola et al. 2003; Blaes et al. 2006). A large class of models are based on relativistic discoseismology, in which the HFQPO frequencies are identified as those of global oscillations modes of the inner accretion discs (see Kato 2001 or Ortega-Rodr\'{i}guez et al. 2008 for reviews of disc oscillations, Lai et al. 2013 for their connection to HFQPOs). Two possible types of oscillations are inertial-gravity modes (or $g$-modes) with vertical structure and inertial-acoustic $p$-modes with no vertical structure. While the $g$-modes exhibit a unique self-trapping property, there is some indication that they may be destroyed by subthermal magnetic fields (Fu \& Lai 2009), comparable to those produced by saturation of the magnetorotational instability, or be destroyed by turbulence (Reynolds \& Miller 2009). Therefore it is unclear whether or not $g$-modes can exist in real discs. However, $p$-modes are only weakly affected by magnetic fields (Fu \& Lai 2011), and their lack of vertical structure makes them insensitive to turbulence.

The types of oscillations which are possible in the disc depend on the properties of its inner edge. Either the gas freely flows into the central black hole, or there exists an inner edge which is impermeable to radial motions, for example due to the presence of a magnetosphere (e.g. Bisnovatyi-Kogan \& Ruzmaikin 1974, 1976; Igumenshchev et al. 2003; Rothstein \& Lovelace 2008; McKinney et al. 2012). In the latter case, there exist ``trapped'' $p$-modes which can propagate between the reflective inner boundary and their inner Lindblad resonances. A trapped $p$-mode can be overstable if it can draw negative energy from the background flow through the corotation resonance. This is possible if the radial derivative of vortensity (vorticity $|\mathbf{\nabla} \times \mathbf{v}|$ divided by surface density) is positive at the corotation radius, which, in the absence of sharp changes in surface density, is not possible in Newtonian discs, but can be achieved with general relativistic (GR) effects (see Tsang \& Lai 2008; Lai \& Tsang 2009; Hor\'{a}k \& Lai 2013). This ``corotational instability'' has been demonstrated in 2D simulations of inviscid discs (Fu \& Lai 2013). 

One of the goals of this paper is to examine the effect of viscosity on trapped $p$-modes. In particular, we are interested in whether viscosity suppresses or enhances their growth. The latter may occur as the result of viscous overstability. We model the disc turbulent viscosity by the standard $\alpha$-ansatz, with different equations of state. The classic theory of viscous overstability describes an axisymmetric instability in which viscosity injects energy drawn from the disc shear into growing oscillations (Kato 1978; Schmit \& Tscharnuter 1995; Schmidt et al. 2001). This effect requires viscous forces to act in phase with oscillations, which is achieved if the dynamic viscosity $\eta$ scales sufficiently steeply with surface density $\Sigma$, i.e., if the parameter $A = \mathrm{d}\ln\eta/\mathrm{d}\ln\Sigma$ is larger than unity. Viscous overstability has been demonstrated in both hydrodynamic (Latter \& Ogilvie 2010) and $N$-body simulations (Rein \& Latter 2013). Extensions of the original axisymmetric theory show that a similar effect can drive non-axisymmetric oscillations (Papaloizou \& Lin 1988; Lyubarskij et al. 1994). Previous analytical calculations, based on local analysis (see Section \ref{sec:theory}), indicate that viscosity can excite disc $p$-modes due to viscous forces acting in phase with the oscillation (e.g. Ortega-Rodr\'{i}guez \& Wagoner 2000; Kato 2001). However, local analysis cannot determine the damping or growth rate of global modes.

In the absence of a reflective boundary, the inner edge of the disc is free-flowing, with gas interior to the ISCO plunging into the central black hole, forming a transonic radial flow. Kato et al. (1988a) found that the sonic point of such flow is unstable to axisymmetric perturbations for sufficiently large viscosities, a mechanism which is distinct from the standard viscous overstability. One-dimensional simulations of transonic discs have shown that viscosity can drive global oscillations at the maximum epicyclic frequency $\kappa_\mathrm{max}$ (Milsom \& Taam 1996; Mao et al. 2009). Two-dimensional and three-dimensional simulations have shown global oscillations at the same frequency (O'Neill et al. 2009), as well as at multiples of innermost stable orbital frequency $\Omega_\mathrm{ISCO}$ (Chan 2009). Another goal of this paper is to study the conditions for producing axisymmetric and various non-axisymmetric modes in transonic discs.

The outline of this paper is as follows. In Section \ref{sec:theory}, we review the theory of viscous overstability, deriving the formula for the growth rate of local non-axisymmetric perturbations due to viscosity. We also provide estimates of the global growth rates of trapped $p$-modes under a pseudo-Newtonian potential. The setup for our numerical experiments is described in Section \ref{sec:setup}. In Section \ref{sec:transonic}, we present the results of simulations of overstable oscillations in a transonic disc with a free inner boundary. We classify several types of global oscillation modes produced in this flow, as well their dependence on the sound speed and viscosity parameter of the disc. In Section \ref{sec:trapped} we examine trapped $p$-modes in a disc with a reflecting boundary, first reproducing semi-analytic results for an inviscid disc, then investigating the effect of viscosity as described by the theory of viscous overstability. As part of this discussion (Section \ref{subsec:newtonian}), we present an unexpected result on overstable modes driven purely by viscosity in the absence of GR effects, whose frequencies are higher than trapped $p$-modes. We summarize our results and discuss their implications in Section \ref{sec:discussion}.

\section{Theory of Viscous Instability}
\label{sec:theory}

\begin{figure*}
	\includegraphics[width=0.49\textwidth]{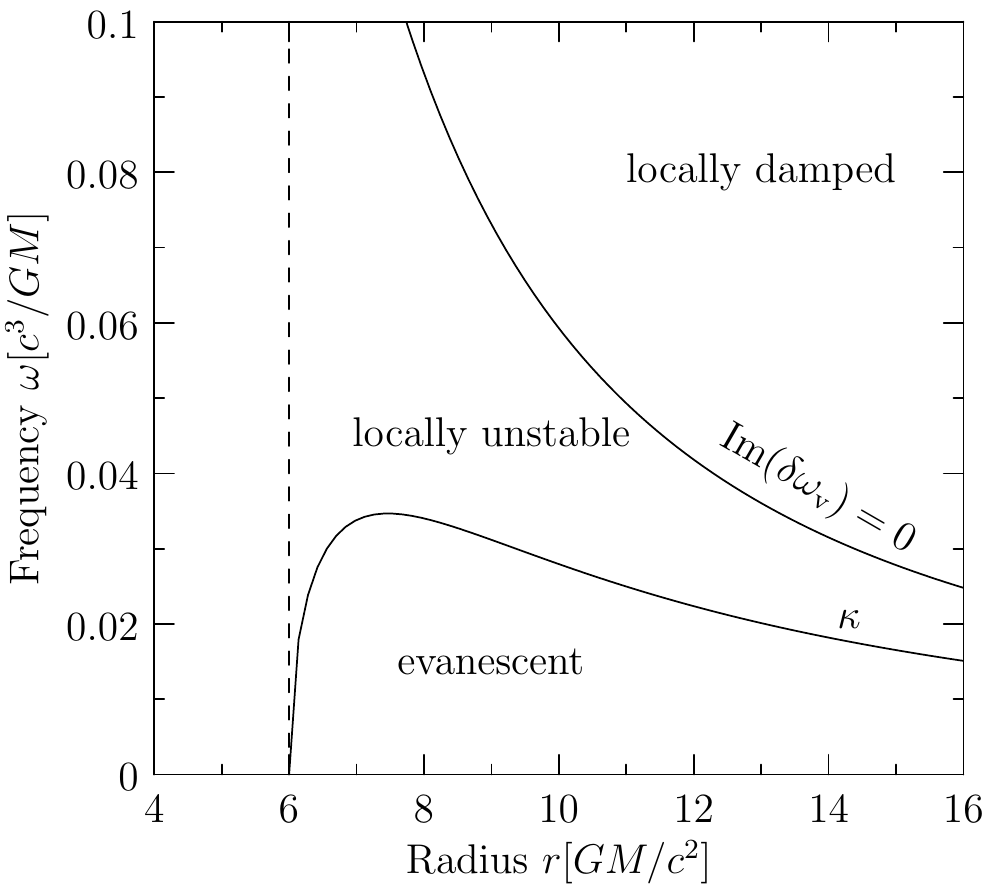}
%	\hfill
	\includegraphics[width=0.49\textwidth]{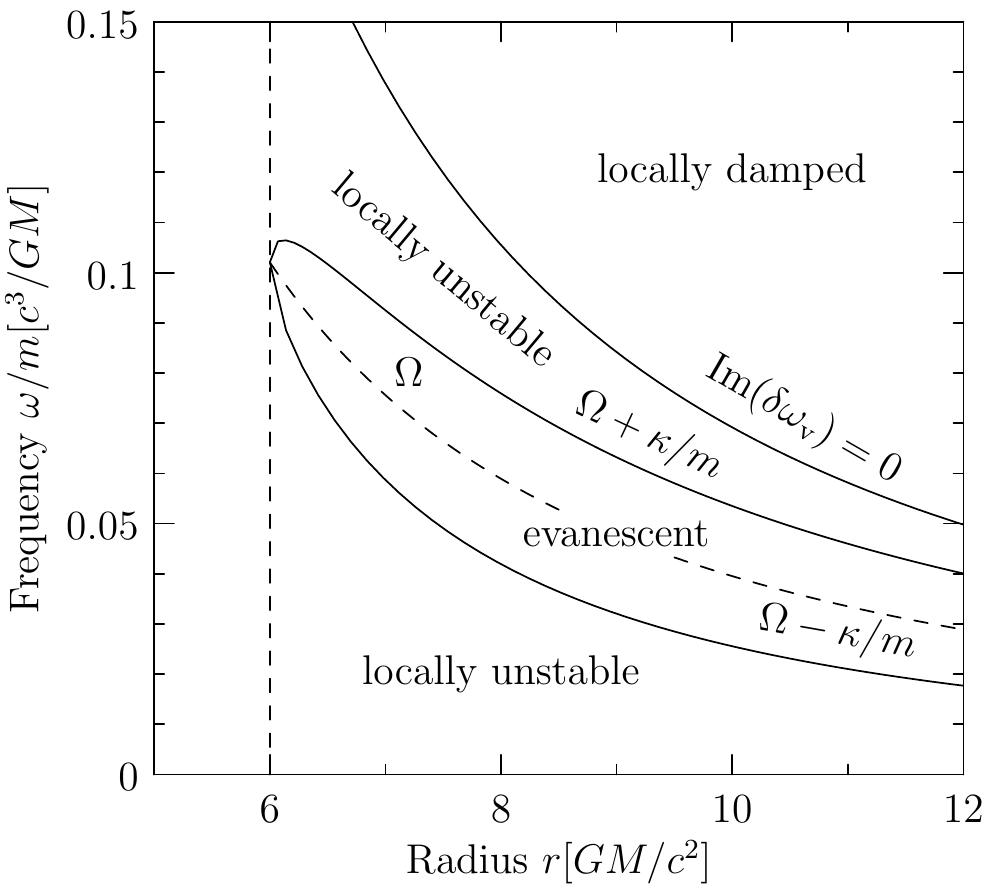}
	\caption{Regions of local viscous stability and instability in the propagation diagram of axisymmetric (left) and non-axisymmetric p-waves. The `evanescent' regions correspond to the cases where $D>0$, the `locally damped' and `locally unstable' regions correspond to positive or negative $\omega_\mathrm{v}$, respectively.}
	\label{fig:propagation-diagrams}
\end{figure*}

The role of viscosity for the stability of thin accretion discs has been studied extensively for several decades. As shown by Kato (1978) in his pioneering work, viscosity can affect the disc stability through both thermal (affecting the energy balance) and dynamical processes (by changing the angular momentum balance). The former process is analogous to so called `$\epsilon$-mechanism' in stellar pulsations. In that case, the instability arises as a consequence of an additional increase in the pressure restoring force, due to viscous heat generation in the compressed phase of the oscillations. In the later case, the instability comes from the mechanical work that is done by the azimuthal component of the viscous force on the fluid elements during the oscillations. To explore this mechanism, it is sufficient to consider barotropic fluid, which we adopt in our work. 

We consider a thin, two-dimensional disc described in polar coordinates $(r,\phi)$ by velocity $\vec{u}=(u_r, u_\phi)$, surface density $\Sigma$ and height integrated pressure $P$ which obey the continuity and Navier-Stokes equations
\begin{align}
	\frac{\partial\Sigma}{\partial t} + \vec{\nabla}\cdot\left(\Sigma\vec{u}\right) &= 0,
	\label{eq:continuity-0}
	\\
	\frac{\partial\vec{u}}{\partial t} + \left(\vec{u}\cdot\vec{\nabla}\right)\vec{u} &=
	-\frac{1}{\Sigma}\vec{\nabla}P - \vec{\nabla}\Phi + 
	\frac{1}{\Sigma}\vec{\nabla}\cdot\boldsymbol{\sigma},
	\label{eq:Navier-Stokes-0}
\end{align}
where $\boldsymbol{\sigma}$ is the viscous stress tensor,
\begin{equation}
	\boldsymbol{\sigma} = \eta\left[\vec{\nabla}\vec{v} + \vec{v}\vec{\nabla} - 
	\frac{2}{3}\left(\vec{\nabla}\cdot\vec{v}\right)\tens{I}\right], 
\end{equation}
$\Phi=\Phi(r)$ is the gravitational potential and $\eta$ is the height integrated dynamic viscosity coefficient that describes momentum transport due to turbulent motion of the fluid. In this section, it is sufficient to assume that both $P$ and $\eta$ are functions of the surface density only, $P=P(\Sigma)$ and $\eta=\eta(\Sigma)$. The particular forms of these functions will be specified later, when the numerical simulations are carried out.

\subsection{Subsonic Part of Disc and Trapped Modes}
In absence of the viscosity ($\eta=0$), the equations (\ref{eq:continuity-0}) and (\ref{eq:Navier-Stokes-0}) admit a stationary axisymmetric solution, that describes a purely rotating flow with the squared angular velocity
\begin{equation}
\label{eq:centrifugal-balance}
	\Omega^2 = \Omega_\mathrm{K}^2 + \frac{c_\mathrm{s}^2}{r^2}\frac{\dd\ln\Sigma}{\dd\ln r},
	\quad
	\Omega_\mathrm{K}^2 = \frac{1}{r}\frac{\dd\Phi}{\dd r},
\end{equation}
where $\Omega_\mathrm{K}$ is the Keplerian angular frequency and $c_\mathrm{s}=(\dd P/\dd\Sigma)^{1/2}$ is the sound speed. Its contribution to the rotational velocity is negligible in thin discs giving nearly Keplerian rotation, $\Omega\approx\Omega_\mathrm{K}$. 

The perturbations to this equilibrium, whose time and azimuthal dependence is of the form of $\exp(\ii m\phi -\ii\omega t)$, obey the wave equation (e.g. Lai \& Tsang 2009),
\begin{equation}
\begin{aligned}
\hat{L}^0 \delta h = &\left[\frac{\dd^2}{\dd r^2} - \left(\frac{\dd}{\dd r}\ln\frac{D}{r\Sigma}\right)\frac{\dd}{\dd r} \right. \\
&\left. + \frac{2m\Omega}{r\tilde{\omega}}\left(\frac{\dd}{\dd r}\ln\frac{D}{\Omega\Sigma}\right) - \frac{m^2}{r^2} - \frac{D}{c_\mathrm{s}^2}\right]\delta h=0
\label{eq:inviscid-wave-eq}
\end{aligned}
\end{equation}
for the enthalpy perturbation $\delta h=\delta P/\Sigma$. Here $\omega$ and $m$ are the angular frequency and azimuthal wavenumber of the perturbation, $\tilde{\omega}=\omega-m\Omega$, $D=\kappa^2 - \tilde{\omega}^2$ and 
\begin{equation}
	\kappa^2 = r\frac{\dd\Omega^2}{\dd r} + 4\Omega^2
\end{equation}
is the squared radial epicyclic frequency. The perturbation of the flow velocity $\delta\vec{u}$ are given as
\begin{equation}
	\delta u_r = \frac{\ii}{D}\left[\tilde{\omega}\frac{\dd}{\dd r} - 
	\frac{2m\Omega}{r}\right]\delta h,
	\quad
	\delta u_\phi = \frac{1}{D}\left[\frac{\kappa^2}{2\Omega}\frac{\dd}{\dd r} - 
	\frac{m\tilde{\omega}}{r}\right]\delta h.
\end{equation}
The operator $\hat{L}^0$ is singular when $D\rightarrow 0$ or $\tilde{\omega}\rightarrow 0$. The first corresponds to the Lindblad resonances and the second to the corotation resonance. While the Lindblad resonances are not real singularities and represent turning points of the waves, the waves may be absorbed at the corotation resonance, which may lead to the instability or damping of the oscillations of the disc.

Introducing the viscosity changes both the equlibrium state and the dynamics of the oscillations. The angular momentum transport causes a slow radial inflow of the matter towards the central black hole in the stationary case. The radial velocity $u_r$ of this inflow can be found by expanding the azimuthal component of equation (\ref{eq:Navier-Stokes-0}) up to the first order in $\eta$,
\begin{equation}
	u_r = \frac{2\Omega}{r^2\kappa^2\Sigma}\frac{\dd}{\dd r}\left(r^3\eta\frac{\dd\Omega}{\dd r}\right).
	\label{eq:radial_drift}
\end{equation}
This expansion however breaks down close to ISCO, where $\kappa\rightarrow 0$ and the flow becomes transonic. The stability of this region of the disc is reviewed later in this section and numerically examined in section~4.

In addition to a small change in the mean flow, the viscosity affects the oscillations through the perturbation of the viscous force $\delta(\vec{\nabla}\cdot\boldsymbol{\sigma})$ that may do a positive or negative work on the waves. Up to the first order in viscosity, the perturbations of the disc are governed by the equation (see Appendix \ref{sec:Appendix1})
\begin{equation}
	\left(\hat{L}^0 + L^1_\mathrm{v} + L^1_\mathrm{in}\right)\delta h = 0,
	\label{eq:visc-wave-eq}
\end{equation}
where $\hat{L}^0$ is the differential operator of the inviscid problem defined in equation (\ref{eq:inviscid-wave-eq}) and $\hat{L}^1_\mathrm{in}$ and $\hat{L}^1_\mathrm{v}$ are its first-order corrections due to the radial inflow in the stationary case and the action of the viscous force:
\begin{align}
	\hat{L}^1_\mathrm{in} &= \ii u_r \frac{\tilde{\omega}}{D}\left[
	\left(1+\frac{\kappa^2}{\tilde{\omega}^2}\right)\frac{\dd^3}{\dd r^3} -
	\frac{D^2}{\tilde{\omega}^2 c_\mathrm{s}^2}\frac{\dd}{\dd r}
	\right],
	\label{eq:L-in}
      \\
	\hat{L}^1_\mathrm{v} &= -\ii\nu\frac{\tilde{\omega}}{D}\left[
	\left(\frac{4}{3} + \frac{\kappa^2}{\tilde{\omega}^2}\right)\frac{\dd^4}{\dd r^4} -
	2qA\frac{D}{\tilde{\omega}^2}\frac{\Omega}{c_\mathrm{s}^2}\frac{\dd^2}{\dd r^2}\right],
	\label{eq:L-v}
\end{align}
where $\nu=\eta/\Sigma$ is the kinematic viscosity.

By substituting the local WKBJ ansatz, $\delta h\propto\exp(\ii k r)$ we may recover the formula of Kato (1978) for the local growth rate of the oscillations due to the action of the viscous force,
\begin{equation}
	\delta\omega_\mathrm{v}(r) = -\ii \nu k^2 \left[\frac{2}{3} + \frac{\Omega^2}{\tilde{\omega}^2}
	\left(\frac{\kappa^2}{2\Omega^2} - qA\right)\right],
	\label{eq:Kato-formula}
\end{equation}
where $q = -\mathrm{d}\ln\Omega/\mathrm{d}\ln r$ and $A = \mathrm{d}\ln\eta/\mathrm{d}\ln\Sigma$. Because $q>0$, the last term reduces the damping and in principle it can lead to an instability. A necessary condition for mode growth [$\mathrm{Im}\left(\delta\omega_\mathrm{v}\right) > 0$] is  positive $A$, i.e. the shear viscosity coefficient $\eta$ increases with increasing density $\Sigma$ on a timescale shorter than the oscillation period ($\sim 1/\Omega$). For a given oscillation frequency, we may separate the regions of local instability or damping of the $p$-waves in the propagation diagram. This is done in Figure~\ref{fig:propagation-diagrams}.

A similar analysis in the case of the operator $\hat{L}_\mathrm{in}$ gives the correction due to the radial inflow in the disc,
\begin{equation}
	\delta\omega_\mathrm{in}(r) = - k u_r.
	\label{eq:inflow-formula}
\end{equation}
The meaning of this result is straightforward; it is the Doppler shift between the stationary observer at a fixed radius and an observer comoving with the fluid. Hence, the inflow only slightly changes the real part of the frequency of the oscillations without affecting their stability. The total frequency change is the sum of the two contributions, $\delta\omega(r)=\delta\omega_\mathrm{v}(r) + \delta\omega_\mathrm{in}(r)$.

Similarly, we may derive a \textit{global} change of the eigenfrequencies of the $p$-modes trapped between two boundaries. A simple calculation presented in Appendix~\ref{sec:Appendix2} leads to the expressions
\begin{equation}
	\delta\omega_\mathrm{v,global} = \frac{\int\delta\omega_\mathrm{v}(r) w(r) \dd r}{\int w(r) \dd r},
	\quad
	\delta\omega_\mathrm{in,global} = \frac{\int\delta\omega_\mathrm{in}(r) w(r) \dd r}{\int w(r) \dd r},
	\quad
	\label{eq:global-growthrates}
\end{equation}
where
\begin{equation}
w(r) = \frac{1}{c_\mathrm{s}}\frac{\tilde{\omega}}{\sqrt{-D}}.
\end{equation}
Hence, the effects of viscosity and radial inflow are just the average of the local rates of equations (\ref{eq:Kato-formula}) and (\ref{eq:inflow-formula}) with the weight function $w(r)$. 

\subsection{Transonic flow}
The studies of the stability of the transonic regions of accretion discs were initiated by the work of Kato et al (1988a). By perturbing equations (\ref{eq:continuity-0}) and (\ref{eq:Navier-Stokes-0}) around a transonic stationary solution describing an isothermal accretion disc, the authors identified two types of unstable axisymmetric perturbations. The first type is essentially a generalization of the inertial-acoustic $p$-waves discussed in the previous subsection to the case of a nonzero radial flow velocity. These modes are trapped between the disc inner edge (corresponding to the sonic radius) and the radius of the inner Lindblad resonance. Consequently, the frequency of these waves is always smaller than $\kappa_\mathrm{max}$, we note however, that this limit is valid only in the case of the axisymmetric perturbations. According to Kato et al. (1988a), as the sonic radius acts only as a partial reflector for the waves, the instability appears only for a sufficiently high viscosity when the viscous driving overcomes the leakage of wave energy through the sonic radius.

The other type of instability appears only in the case of a transonic flow and represents a standing wave pattern localized around the sonic radius that grows exponentially with time. Kato et al. (1988a) assumed so-called conventional or `$\alpha p$'-type viscosity, in which the $r\phi$-component of the viscous stress tensor is directly proportional to the pressure, $\sigma_{r\phi} = \alpha p$ (contrary to our `diffusive' prescription for the viscosity, that relates pressure to the shear viscosity coefficient $\eta$). They found that the instability occurs only for high enough values of $\alpha$, when $\alpha\geq u^\prime_\mathrm{c}/\Omega_\mathrm{c}$, with $u^\prime_\mathrm{c}$ and $\Omega_\mathrm{c}$ being a radial velocity gradient and orbital frequency of the flow at the sonic radius. They also noted that the same condition is satisfied when the flow changes from being subsonic to supersonic by passing through the nodal critical point. They speculated that the appearance of this second type of instability is directly related to the topology of the flow at the sonic radius. This idea was supported in subsequent studies: Kato et al. (1988b) relaxed the assumption of isothermal flow by including the energy balance between viscous heating and radiative cooling and found that the instability criterion coincides with the one for the nodal topology of the sonic point. Later on, Kato et al (1993) examined a stability of the isothermal accretion flows with the diffusive form of the viscosity similar to our work. In that case the sonic point is always of the saddle type and consequently the authors found that it is stable against this type of perturbation.

Based on these results we expect that possible instabilities of our flow arise only due to propagating acoustic waves mentioned in the beginning of this subsection. In the following sections, we will examine numerically the conditions for their growth as well as the spatial structure of the unstable modes.

\section{Numerical Setup}
\label{sec:setup}
As in the previous section, we consider a thin disc described by velocity $\mathbf{u} = \left(u_r, u_\phi\right)$, surface density $\Sigma$ and pressure $P$ which obey the Navier-Stokes equations. It is subject to the ``pseudo-Newtonian'' (Paczy\'{n}ski \& Wiita 1980) gravitational potential 
\be
\Phi = -\frac{GM}{r-r_\mathrm{s}},
\ee
where $M$ is the mass of the central black hole and $r_{\mathrm{s}} = 2GM/c^2$ is its Schwarzschild radius. This potential mimics GR effects, having a Keplerian orbital frequency and radial epicyclic frequency given by
\be
\Omega_\mathrm{K} = \sqrt{\frac{GM}{r^3}}\left(\frac{r}{r-r_\mathrm{s}}\right), \quad \kappa = \Omega_\mathrm{K}\sqrt{\frac{r-3r_\mathrm{s}}{r-r_\mathrm{s}}}.
\ee
There is an innermost stable circular orbit (ISCO) defined by $\kappa^2\left(r_\mathrm{ISCO}\right) = 0$, and a radius at which the epicyclic frequency peaks, $\kappa\left(r_\mathrm{max}\right) = \kappa_\mathrm{max} = \left(9-5\sqrt{3}\right)\Omega_{\mathrm{ISCO}}$, which are located at $r_\mathrm{ISCO} = 3r_{\mathrm{s}}$ and $r_{\mathrm{max}} = \left(2 + \sqrt{3}\right)r_{\mathrm{s}}$, respectively.

We adopt numerical units such that
\be
r_\mathrm{ISCO} = \Omega_\mathrm{ISCO} = 1,
\ee
and define an ``orbit'' as one orbital period at $r_{\mathrm{ISCO}}$ (equal to $2\pi$ in numerical units). We use a polytropic equation of state $P = K\Sigma^\Gamma$, which has the corresponding sound speed $c_\mathrm{s} = \sqrt{\Gamma P/\Sigma}$. We define the parameter $c_{\mathrm{s}0} = c_\mathrm{s}\left(r_\mathrm{ISCO}\right)$, whose value we will refer to rather than that of the constant $K$, to which it is directly related. We prescribe the kinematic viscosity of the fluid as
\be
\nu = \alpha c_\mathrm{s} H, 
\ee
where $H = c_\mathrm{s}/\Omega_\mathrm{K}$ is the disc scale height (Shakura \& Sunyaev 1973). Note that $\nu$ is proportional to $c_\mathrm{s}^2$, so that $\nu\left(\Sigma\right) \propto \Sigma^{\Gamma - 1}$ or $\eta\left(\Sigma\right) \propto \Sigma^\Gamma$, giving $A = \mathrm{d}\ln\eta / \mathrm{d}\ln\Sigma = \Gamma$. Therefore each simulation is defined by three parameters: $c_{\mathrm{s}0}$, $\Gamma$ and $\alpha$. 

The outer boundary is always located at $r = 4$, at which we fix $\Sigma$ and $u_\phi$ at their initial values and impose $\partial u_r/\partial r = 0$. Adjacent to the outer boundary, we implement a wave damping zone of width $1/3$ (i.e., between $r = 11/3$ and $r = 4$) in which a damping force (per unit mass), given by
\be
f_\mathrm{damp} = -\frac{\mathbf{u}-\mathbf{u}_0}{\tau}R\left(r\right)
\ee
is applied (de Val-Borro et al. 2006). Here $\mathbf{u}_0 = \left(r\Omega_0,0\right)$, $\tau$ is a damping timescale equal to the orbital period at the outer boundary, and $R\left(r\right)$ is a parabolic function, which is equal to unity at $r = 4$ and reduces to zero at $r = 11/3$. The damping zone minimizes reflection of waves from the outer boundary and mimics an outgoing wave boundary condition used in linear analysis of disc modes (e.g., Lai \& Tsang 2009). We simulate two distinct physical setups which correspond to two different inner boundary locations and their corresponding boundary conditions, choosing either a free boundary at $r = 2/3$ (Section \ref{sec:transonic}) or a reflecting boundary at $r = 1$ (Section \ref{sec:trapped}). The details and significance of these boundary conditions are discussed in more detail in their corresponding sections. The initial conditions of the simulations always have $u_r = 0$ everywhere, but the initial surface density profile $\Sigma\left(r\right)$ varies in the different sections of this paper. The orbital velocity $u_\phi$ is always chosen to be initially in centrifugal balance (equation \ref{eq:centrifugal-balance}) given the surface density profile.

The Navier-Stokes equations are solved using the PLUTO code (Mignone et al. 2007) with third-order Runge-Kutta time stepping, parabolic reconstruction and a Roe solver on static polar grid with uniform spacing in both directions. The standard resolution of the simulations is $N_r \times N_\phi = 1024 \times 256$, so that the radial direction has a spatial resolution of at least $300$ per unit $r$ (this varies slightly with the location of the inner boundary), and modes with $m \leq 4$ are captured with at least $64$ zones per azimuthal wavelength. The sensitivity of our numerical results to the chosen resolution is addressed seperately for each major results section (Sections \ref{sec:transonic} and \ref{sec:trapped}).

\section{Overstable Global Oscillations in Transonic Discs}
\label{sec:transonic}

\begin{table*}
\begin{center}
\begin{tabular}{|c|c|c|c|c|c|c|c|}
\hline
\hline

Label & $c_{\mathrm{s}0}$ & $\alpha$ & $r_\mathrm{c}$ & $u'_\mathrm{c}/\Omega_\mathrm{c}$ & $\kappa_\mathrm{max}$ & $m\Omega_\mathrm{ISCO}$ & $m\Omega_\mathrm{ISCO} \pm \kappa_\mathrm{max}$ \\ \hline
(a) & $0.01$ & $0.05$ & $0.99$ & $0.150$ & No & No & No \\
(b) & $0.01$ & $0.10$ & $1.02$ & $0.120$ & No & $4$ & No \\
(c) & $0.01$ & $0.25$ & $1.12$ & $0.087$ & Linear & $2$ & Yes \\
(d) & $0.01$ & $0.50$ & $1.30$ & $0.082$ & Nonlinear & $1$ & Yes \\
(e) & $0.02$ & $0.10$ & $1.01$ & $0.170$ & No & No & No \\
(f) & $0.02$ & $0.25$ & $1.17$ & $0.109$ & No & $1$ & Yes \\
(g) & $0.02$ & $0.50$ & $1.69$ & $0.116$ & Nonlinear & No & No \\
(h) & $0.05$ & $0.25$ & $1.12$ & $0.197$ & No & No & No \\
(i) & $0.05$ & $0.50$ & $1.57$ & $0.446$ & Nonlinear & No & No \\ \hline

\end{tabular}
\end{center}
\caption{Summary of transonic disc simulations. The first three columns give the name of the run and the value of parameters $c_{\mathrm{s}0}$ and $\alpha$. The next two columns give the numerically measured location of the sonic point $r_\mathrm{c}$ and its dimensionless stability quantity $u'_\mathrm{c}/\Omega_\mathrm{c}$. The last three columns indicate whether or not overstable oscillations of the three types described in Section \ref{subsec:ts_classification} are present. For the $\kappa_\mathrm{max}$ type, we also indicate whether or not they are linear or nonlinear, and if the $m\Omega_\mathrm{ISCO}$ type are present, we indicate which azimuthal number $m$ is dominant in the power spectrum.}
\label{tab:transonic_summary}
\end{table*}

\begin{figure}
\begin{center}
\includegraphics[width=0.49\textwidth]{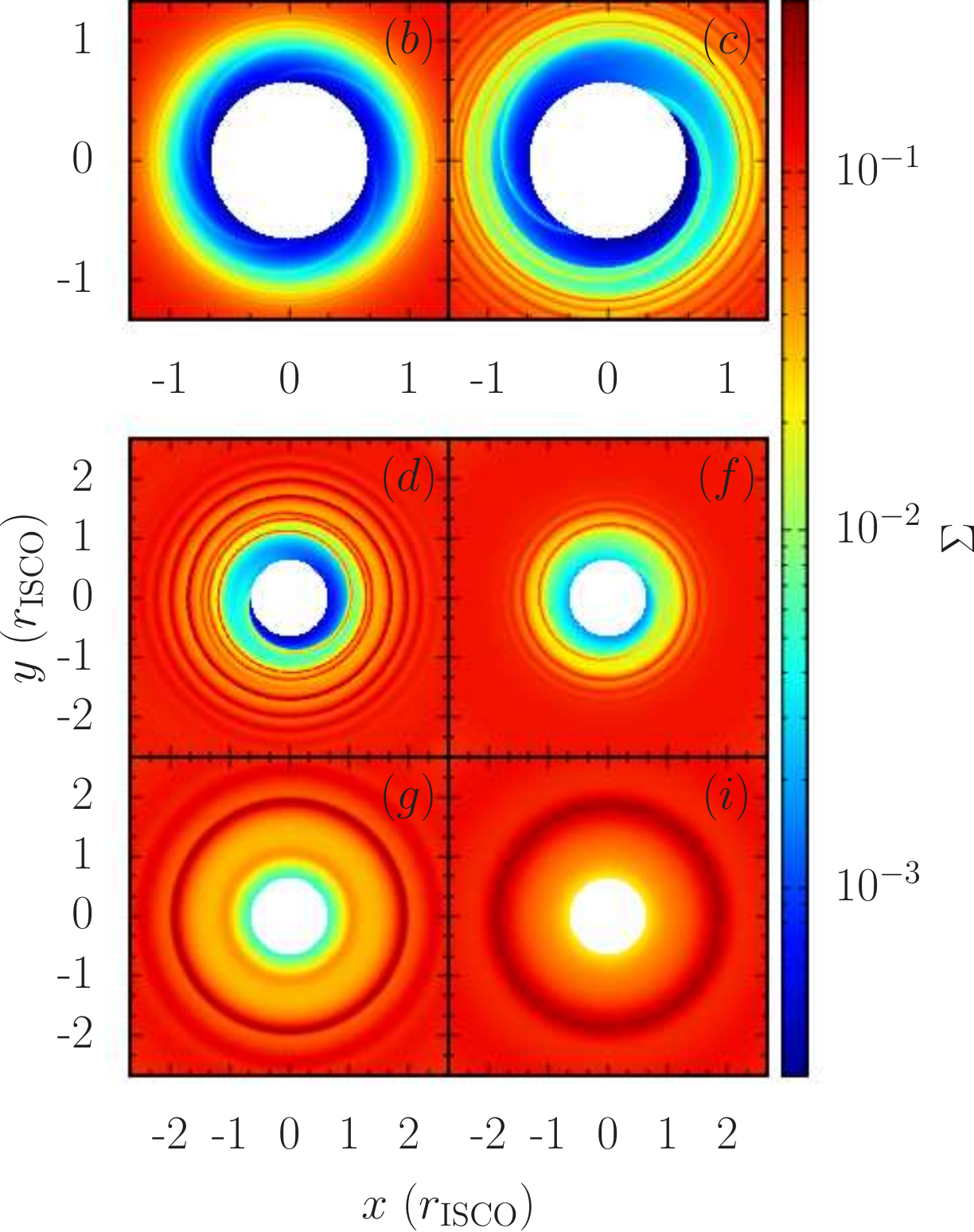}
\caption{Surface density at $t = 100$ for runs with significant time variability. The parameters corresponding to the alphabetical labels are given in Table \ref{tab:transonic_summary}. Various features can be seen, including four-armed, two-armed and one-armed spirals, as well as axisymmetric rings. The top two panels are zoomed in relative to the other panels to show finer detail of the non-axisymmetric structure close to the inner edge.}
\label{fig:snapshot_ts}
\end{center}
\end{figure}

\begin{figure}
\begin{center}
\includegraphics[width=0.49\textwidth]{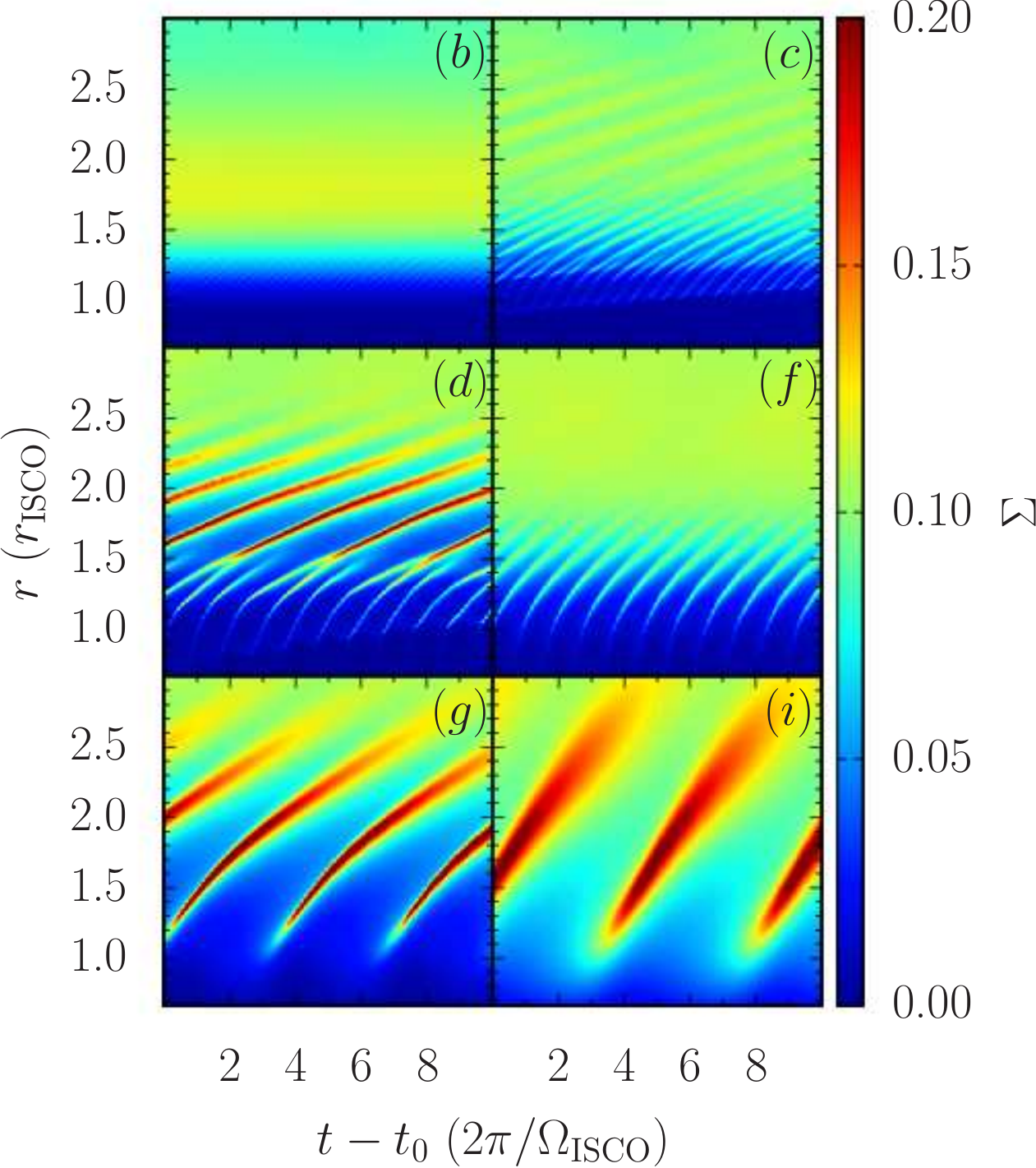}
\caption{Space-time diagrams of surface density $\Sigma$ at $\phi = 0$, demonstrating the structure of the spatial variabilities and temporal variabilities of the oscillations. Time is shown relative to the reference time $t_0 = 90$. We can see that waves with frequencies close to multiple of $\Omega_\mathrm{ISCO}$ (if they are present) do not propagate at radii much larger than $r_\mathrm{ISCO}$, beyond which only axisymmetric oscillations with frequency approximately $\kappa_\mathrm{max}$ propagate.}
\label{fig:prop_ts}
\end{center}
\end{figure}

\begin{figure}
\begin{center}
\includegraphics[width=0.49\textwidth]{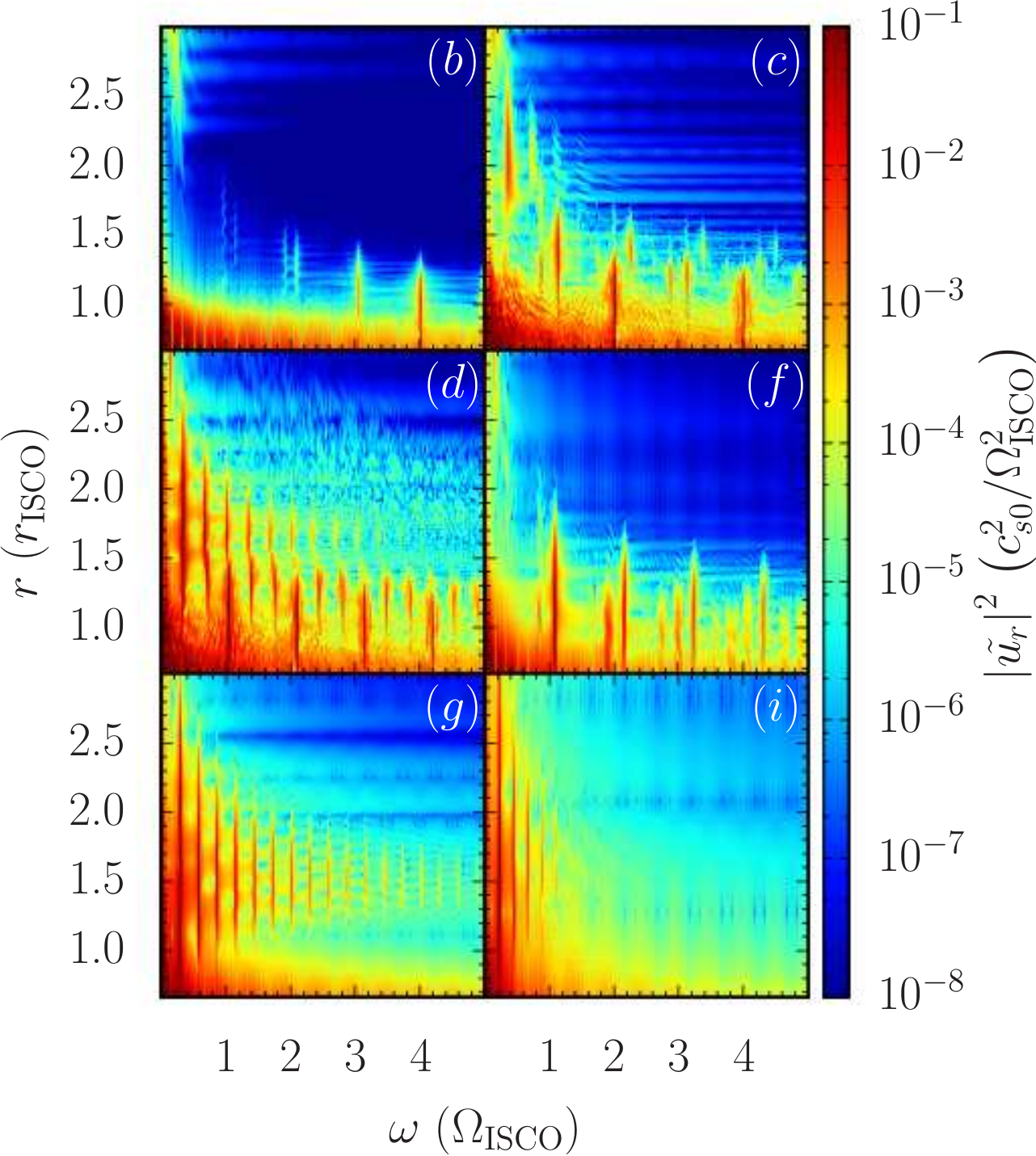}
\caption{The power spectrum $|\tilde{u}_r|^2$ over a range of disc radii. Global modes with power in discrete frequencies which are coherent across a range of $r$ are present in all cases. They can be seperated into modes with frequencies close to multiples of $\Omega_\mathrm{ISCO}$ (and their nonlinear splittings) with power concentrated near $r_\mathrm{ISCO}$, and those with frequencies close to multiples of $\kappa_\mathrm{max}$ with power concentrated at larger radii. These correspond respectively to the spirals and axisymmetric rings seen in Figure \ref{fig:snapshot_ts}.}
\label{fig:psd_ts}
\end{center}
\end{figure}

\begin{figure}
\begin{center}
\includegraphics[width=0.49\textwidth]{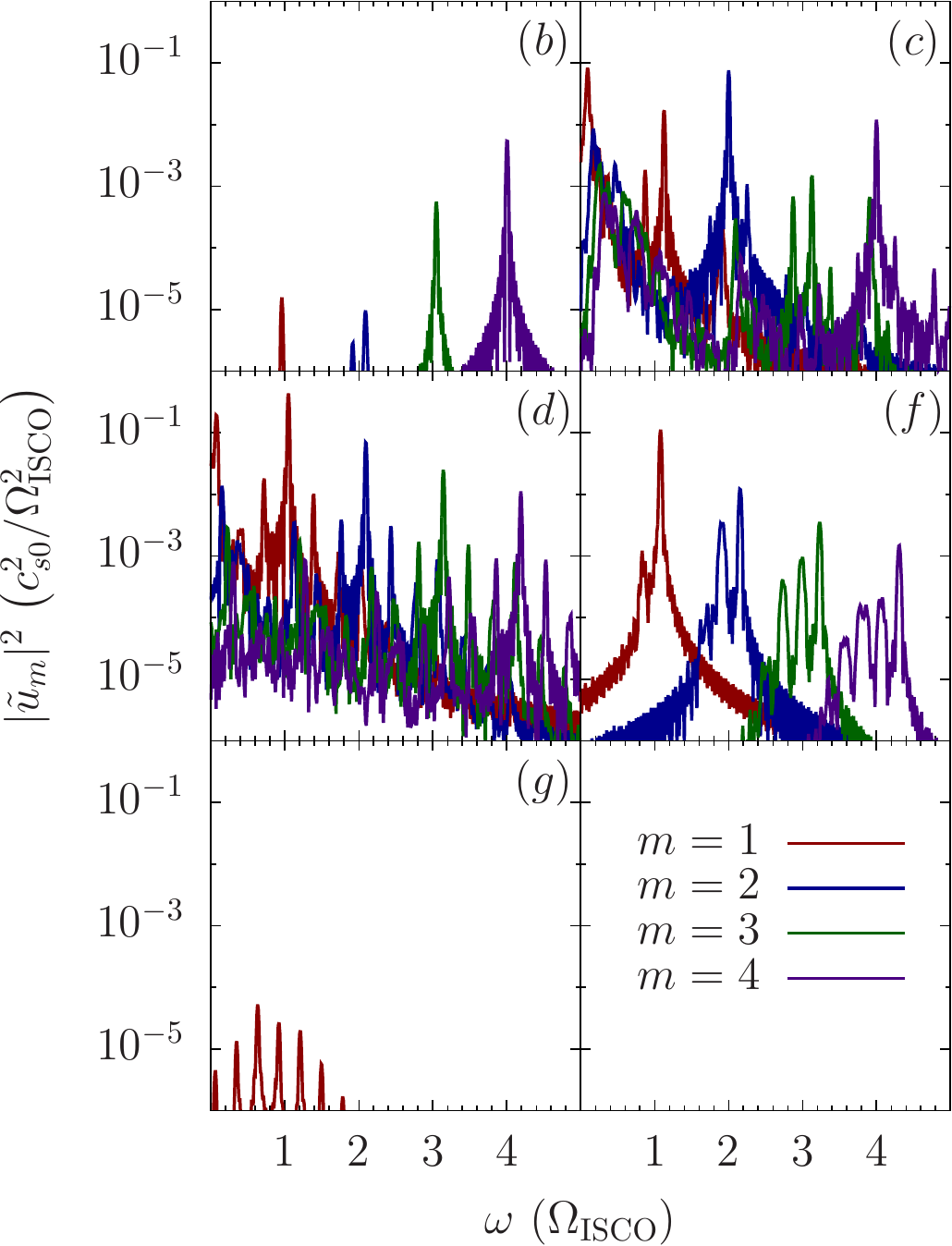}
\caption{Power spectrum delineated by azimuthal number $m$, given by the quantity $|\tilde{u}_m|^2$ defined in eq. (\ref{eq:mfreq}).}
\label{fig:mfreq_ts}
\end{center}
\end{figure}

For our transonic disc simulations, the inner boundary is placed at $r_\mathrm{in} = 2/3$, at which an outflow boundary condition given by
\be
\frac{\partial \Sigma}{\partial r} = \frac{\partial \mathbf{u}}{\partial r} = 0
\ee
is imposed. We choose as our initial condition
\be
\Sigma \propto \frac{r^{-\frac{1}{2}}}{r-r_\mathrm{s}} \left[1 - \left(\frac{r}{r_\mathrm{in}}\right)^{-\frac{1}{2}}\right],
\ee
which is an approximate steady-state surface density profile consistent with constant $\dot{M}$ and vanishing torque at $r_\mathrm{in}$, but it is not a solution for the transonic flow region interior to $r_\mathrm{ISCO}$. From these initial conditions, the disc is evolved for $100$ orbits. We wait for it to settle into a quasi-steady state before analyzing its variabilities. The parameters for various runs and their corresponding key results are summarized in Table \ref{tab:transonic_summary}. Run $(a)$, which has the longest viscous timescale, is the slowest to reach a quasi-steady state, taking $\sim 60$ orbits (however, this is much less than the viscous timescale). Therefore we perform all analysis in the last $30$ orbits of each run, at which point all have reached a quasi-steady state. Since we ignore the initial transient phase, we do not attempt to characterize the growth process or linear behavior of any observed oscillations. Instead, we focus on the properties of unstable oscillations which have saturated at some (often large) amplitudes and which may have become significantly nonlinear. 

We use the following formalism to analyze the variabilities of the quasi-steady, nonlinear phase of the simulations. We define the power spectrum $|\tilde{u}|^2$, where
\be
\tilde{u}\left(r,\omega\right) = \frac{1}{2\pi}\int_{t_1}^{t_2} u_r\left(r,0,t\right) e^{-i \omega t} \mathrm{d}t
\ee
is the Fourier transformed radial velocity and $\left(t_1,t_2\right) = \left(70,100\right)$. The power spectrum gives a measure of the strength of the oscillations, along with their frequencies and locations in the disc where they are most visible. We determine the azimuthal number $m$ associated with each frequency by decomposing the radial velocity into components proportional to $\exp\left(im\phi-i\omega t\right)$. Since, as we will demonstrate, the oscillations are global (their frequencies are coherent across a range of radii), we average this amplitude over all $r$. The resulting quantity,
\be
\begin{aligned}
\tilde{u}_m\left(\omega\right) &= \frac{1}{4\pi^2\left(r_{\mathrm{out}}-r_{\mathrm{in}}\right)} \\ 
&\times \int_{r_\mathrm{in}}^{r_{\mathrm{out}}}\int_{t_1}^{t_2} \int_0^{2\pi} u_r\left(r,\phi,t\right) e^{-\left(im\phi-i\omega t\right)} \mathrm{d}\phi \mathrm{d}t \mathrm{d}r,
\end{aligned}
\label{eq:mfreq}
\ee
specifies the global power spectrum delineated by $m$. The spatial and temporal variability and propagation properties of the oscillations are examined by plotting the surface density $\Sigma$ at a fixed azimuthal angle ($\phi = 0$) as a function of radius $r$ and time $t$.

For this section we restrict the equation of state to only the isothermal case, $\Gamma = 1$, so that each run is characterized only by the parameters $c_{\mathrm{s}0}$ and $\alpha$. The parameters of the nine numerical runs (see Table \ref{tab:transonic_summary}) were chosen because they lead to a variety of behaviors, producing overstable oscillations of several types (to be described below) which are present or absent in various combinations. For each sound speed $c_{\mathrm{s}0}$, there is a threshold value of the viscosity parameter $\alpha$ below which no overstable oscillations are observed within the duration of the run ($100$ orbits). Three of our runs, ($a)$, $(e)$ and $(h)$, fall into this regime, leaving six runs which show significant variabilities. The subsequent discussion focuses only on these six interesting cases.

Figure \ref{fig:snapshot_ts} shows snapshots of the disc surface density $\Sigma$ at $t = 100$ for the runs that exhibit overstable oscillations. Figure \ref{fig:prop_ts} shows $\Sigma$ at $\phi = 0$ as function of time and radius. Together they illustrate the spatial structure and wave propagation properties of the modes. In runs $(b)$, $(c)$, $(d)$ and $(i)$, non-axisymmetric waves with frequencies approximately multiples of $\Omega_\mathrm{ISCO}$ propagate at small radii ($r \lesssim 1.2$ for $c_{\mathrm{s}0} = 0.01$, $r \lesssim 1.7$ for $c_{\mathrm{s}0} = 0.02$). In runs $(c)$, $(d)$, $(g)$ and $(i)$, axisymmetric waves with lower frequencies propagate at larger radii. In cases in which both types of waves are present, there is a merging or winding-up of non-axisymmetric waves into axisymmetric ones at intermediate radii. 

Figure \ref{fig:psd_ts} depicts the power spectra, for a range of radial locations in the disc, of the runs which exhibit overstable oscillations. In all cases, power is sharply concentrated in discrete frequencies and present across a large range of $r$, manifesting as narrow vertical strips in Figure \ref{fig:psd_ts}. We therefore interpret these as frequencies of coherent global modes, rather than local oscillations whose properties vary with $r$. Since they are global, with frequencies independent of location, we can justify the radial integration in equation (\ref{eq:mfreq}), which allows us to analyze their relative strengths globally, rather than locally. The $m$-delineated power spectra are shown in Figure \ref{fig:mfreq_ts}. Note that the absolute scale of the quantity $|\tilde{u}_m|^2$ does not have a direct physical meaning, as it depends on the radial integration range (which we choose as the entire computational domain, whose size is arbitrary). However the relative amplitudes, both between different values of $m$ and across multiple panels in Figure \ref{fig:mfreq_ts}, give a meaningful comparison of the power in the various modes. From these we see that run $(b)$, $(c)$, $(d)$ and $(f)$ are dominated by $m = 4$, $m = 2$, $m = 1$ and $m = 1$ modes, respectively. Runs $(g)$ and $(i)$ have very little power in non-axisymmetric modes, with the majority of their power in axisymmetric ($m = 0$) modes, whose corresponding $|\tilde{u}_m|$ is not shown.

\subsection{Classification of Oscillations}
\label{subsec:ts_classification}
We classify the observed oscillation modes into three types. First, there is a ``$\kappa_\mathrm{max}$'' type, which is axisymmetric ($m = 0$) and has a frequency approximately equal to (for small $c_{\mathrm{s}0}$) or somewhat smaller than (for large $c_{\mathrm{s}0}$) the maximum epicyclic frequency, $\kappa_\mathrm{max} = 0.34$. Under various conditions, this mode is either linear or nonlinear. In the former case, its wave function is smooth and has a small amplitude, and only the fundamental frequency is present in the power spectrum. In the latter case, the wave function is sharp and large in amplitude, and many overtones (integer multiples of $\kappa_\mathrm{max}$) are present in the power spectrum. This behavior is analogous to that of a nonlinear oscillator with natural frequency $\kappa_\mathrm{max}$. As the amplitude of the oscillation becomes large, nonlinear effects introduce resonances at multiples of the natural frequency, and oscillations at these frequencies can also be driven to large amplitudes. The result is the excitation of a series of harmonics of the frequency $\kappa_\mathrm{max}$. In the most extreme examples of this nonlinearity [e.g. runs $(d)$ and $(g)$], the power density (see Figure \ref{fig:psd_ts}) in the first three to four overtones is comparable to (less than, but within an order of magnitude of) that of the fundamental frequency, and up to ten overtone frequencies can be identified [run $(g)$].

The second type of mode is the ``$m \Omega_\mathrm{ISCO}$'' type, which is non-axisymmetric with azimuthal number $m > 0$ and frequency $\omega \approx m \Omega_\mathrm{ISCO}$. These always appear in a series of successive $m$, and therefore represent a harmonic series of the fundamental frequency (similar to the nonlinear effect seen in the $\kappa_\mathrm{max}$ modes), which is simply $\Omega_\mathrm{ISCO}$. However, the $m$ with the largest power is not always $1$, we find cases in which it is $2$ or $4$. This manifests as a sharp, nonlinear $m$-armed spiral in the disc.

The third type of mode is the ``$m\Omega_\mathrm{ISCO} \pm \kappa_\mathrm{max}$'' type, which has azimuthal number $m > 0$ and frequency approximately $m \Omega_\mathrm{ISCO} \pm \epsilon$, where $\epsilon$ is close to, but sometimes less than, $\kappa_\mathrm{max}$. These are prominently present for moderate values of $\alpha$ when $c_{\mathrm{s}0}$ is small [runs $(c)$ and $(d)$], moderately present for a similar $\alpha$ at a larger $c_{\mathrm{s}0}$ [run $(f)$] and entirely absent for large $\alpha$ or large $c_{\mathrm{s}0}$ [runs $(g)$ and $(i)$]. We suggest that these modes are a result of a splitting of the $m \Omega_\mathrm{ISCO}$ modes due to a nonlinear coupling with the $\kappa_\mathrm{max}$ modes, as these are the most fundamental modes associated with the two ``special'' locations in the disc, $r_\mathrm{ISCO}$ and $r_\mathrm{max}$.

\subsection{Dependence on Parameters}
From our runs, we can extrapolate the dependence on the parameters $c_{\mathrm{s}0}$ and $\alpha$ in determining power spectrum of overstable modes. For a given sound speed $c_{\mathrm{s}0}$, we imagine slowly increasing the value of $\alpha$, starting from a very small value. A sufficiently small $\alpha$ produces no overstable oscillations, leading to a steady-state axisymmetric flow with no variabilities. As $\alpha$ increases, there is a threshold value at which $m\Omega_\mathrm{ISCO}$ type modes begin to appear. This threshold value of $\alpha$ is larger for larger $c_{\mathrm{s}0}$. For example, the threshold is $\alpha \gtrsim 0.05$ for $c_{\mathrm{s}0} = 0.01$ and $\alpha \gtrsim 0.1$ for $c_{\mathrm{s}0} = 0.02$. Once this threshold is reached, the dominant $m\Omega_\mathrm{ISCO}$ mode initially involves large $m$, such as $m = 4$ seen in run $(b)$. As $\alpha$ is further increased, the dominant $m$ of these modes becomes smaller, for example $m = 2$ in run $(c)$ and $m = 1$ in run $(d)$. As $\alpha$ increases further, beyond a second threshold value, axisymmetric $\kappa_\mathrm{max}$ modes become present, while the $m\Omega_\mathrm{ISCO}$ modes may still be present with comparable power, as in run $(d)$, or may be suppressed, such that only $\kappa_\mathrm{max}$ modes are present, as in runs $(g)$ and $(i)$. In between the two threshold values of $\alpha$, it is possible for the $m\Omega_\mathrm{ISCO} \pm \kappa_\mathrm{max}$ modes to be present, sometimes in the absence of strong $\kappa_\mathrm{max}$ modes, such as in runs $(c)$ and $(f)$, probably due to a nonlinear coupling between the two other types of modes.

As reviewed in Section \ref{sec:theory}, Kato et al. (1988a) derived a criterion a for a viscous pulsational instability at the sonic point $r_\mathrm{c}$, defined by $u_r\left(r_\mathrm{c}\right) = c_\mathrm{s}$, of a transonic flow. According to this criterion, the sonic point is unstable if $\alpha > u'_\mathrm{c}/\Omega_\mathrm{c}$, where $u_\mathrm{c}$ and $\Omega_\mathrm{c}$ are the radial velocity and angular velocity at the critical point and the prime denotes a radial derivative. Note that these quantities themselves depend on $\alpha$ (and $c_{\mathrm{s}0}$) in a self-consistent solution of the flow. While this instability criterion may not be exactly applicable in our simulations (due to our `diffusive' rather than `$\alpha P$' viscosity prescription), we find it is still a useful diagnostic. Using azimuthally-averaged and time-averaged profiles of $u_r$ and $u_\phi$, we measure the location of the sonic point $r_\mathrm{c}$ and the value of $u'_\mathrm{c}/\Omega_\mathrm{c}$ for each run, which are presented in Table \ref{tab:transonic_summary}. From these we can see that the Kato et al. criterion successfully indicates whether or not overstable oscillations are found, with the exception of run $(b)$, for which a weak $m = 4$ mode is found even though the criterion is not strictly satisfied.

The results of our simulations of transonic discs can be most directly compared to those of Chan (2009), who also examined the role of viscosity in height-integrated discs (with a more general equation of state), and found evidence for viscous excitation of global axisymmetric and non-axisymmetric $p$-modes of the types found in our simulations. Our work extends Chan's analysis by delineating the different regimes (ranges of $c_{\mathrm{s}0}$ and $\alpha$) in which particular combinations of these modes are excited, and highlights the possibility of extreme nonlinearities and the existence of the mixed $m\Omega_\mathrm{ISCO} \pm \kappa_\mathrm{max}$ type modes. The driving of the $\kappa_\mathrm{max}$ modes at large $\alpha$ is consistent with the 1D (height-integrated and axisymmetric) simulations of Milsom \& Taam (1996) and Mao et al. (2009), as well as the 2D (axisymmetric) simulations of O'Neill et al. Note that the aforementioned studies are unable to capture the global spiral modes seen in our simulations due to their assumptions of axisymmetry.

\subsection{Resolution Study}
\label{subsec:ts_resolution}

\begin{figure}
\begin{center}
\includegraphics[width=0.49\textwidth]{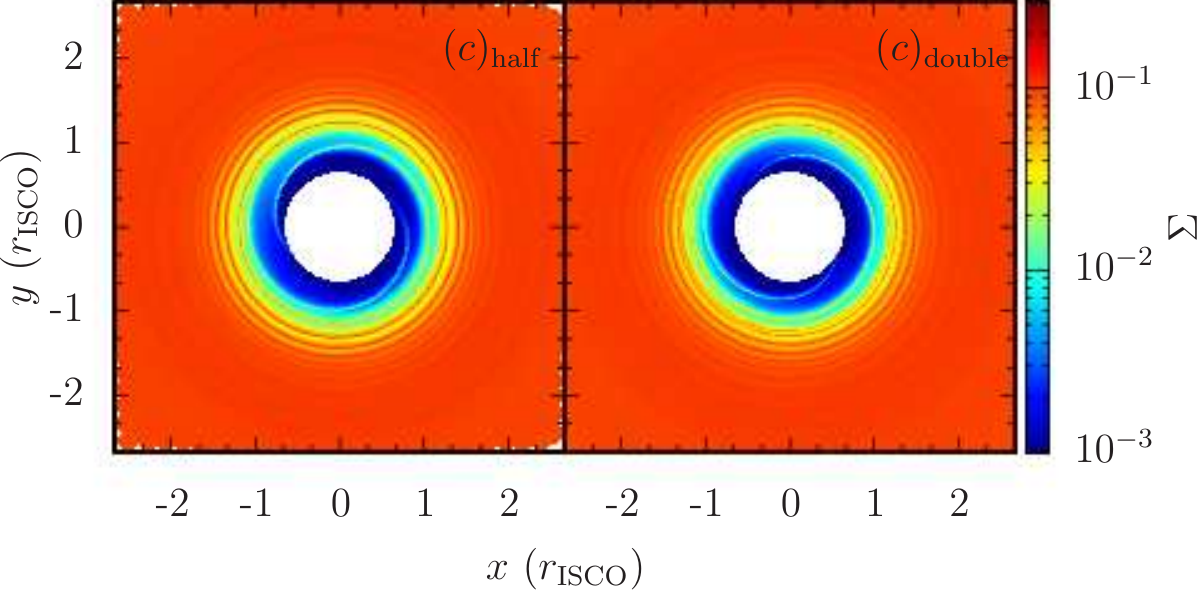}
\includegraphics[width=0.49\textwidth]{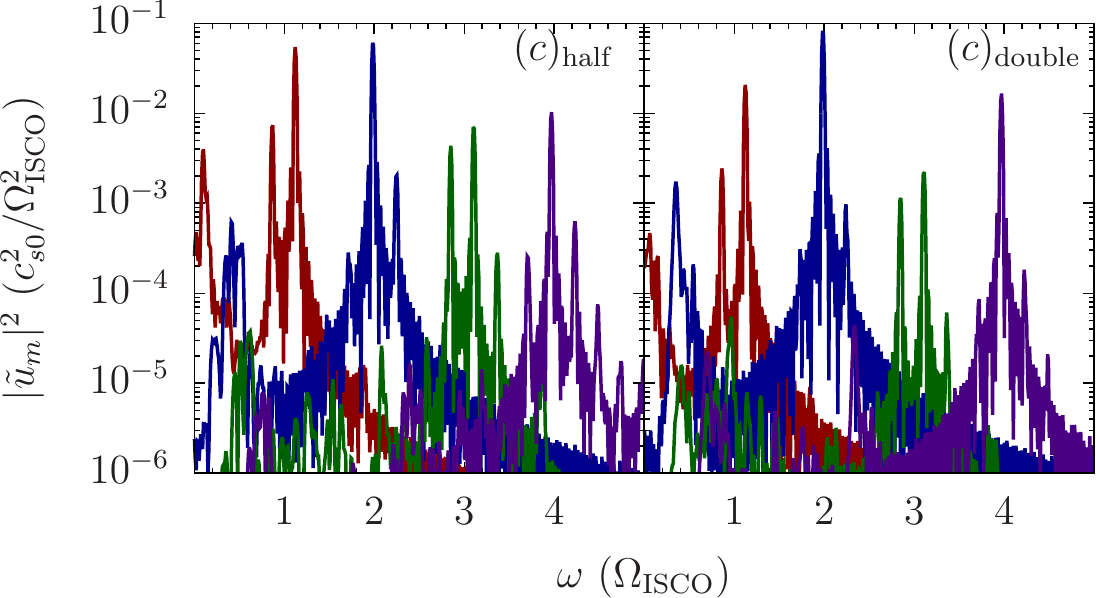}
\caption{Surface density at $t = 100$ and power spectrum of resolution test runs $(c)_\mathrm{half}$ and $(c)_\mathrm{double}$ (compare to Figures \ref{fig:snapshot_ts} and \ref{fig:mfreq_ts}).}
\label{fig:ts_resolution}
\end{center}
\end{figure}

In order to determine the robustness and dependence on grid resolution of the features seen in our transonic disc simulations, we perform run $(c)$ with both half the standard resolution in both dimensions, $N_r \times N_\phi = 512 \times 128$, denoted $(c)_\mathrm{half}$ and double the standard resolution $N_r \times N_\phi = 2048 \times 512$, denoted $(c)_\mathrm{double}$. The surface density at $t = 100$ and the $m$-delineated power spectra for these runs are shown in Figure \ref{fig:ts_resolution}. It can be seen from these that the runs are remarkably similar to one another, as well as to the standard resolution run, therefore we conclude that the observed features are not sensitive to resolution and that the standard resolution provides sufficient convergence. Since run $(c)$ has the largest Reynolds number $\mathrm{Re} = \alpha^{-1}c_{\mathrm{s}0}^{-2}$ of the runs which exhibit all three types of oscillations, we conclude that our other runs are at least as well resolved as $(c)$ under the standard resolution.

\section{Overstable Global Oscillations in Truncated Discs}
\label{sec:trapped}

\begin{table*}
\begin{center}
\begin{tabular}{|c|c|c|c|c|c|c|}
\hline
\hline

& $m$ & $\left(\omega_\mathrm{r}, \omega_\mathrm{i}\right)_{\mathrm{semi-analytic}}$ & $\omega_\mathrm{r}$ & $\omega_{\mathrm{i},\alpha=0}$ & $\omega_{\mathrm{i},\alpha=0.01}$ & $\omega_{\mathrm{i},\alpha=0.05}$ \\ \hline

$\boldsymbol{c_{\mathrm{s}0} = 0.02}$, $\boldsymbol{\Gamma = 1}$
& $2$ & $1.78, 0.0422$ & $1.78$ & $0.0453$ & $0.0314$ & $0.0266$ \\
& $3$ & $2.69, 0.0528$ & $2.69$ & $0.0579$ & $0.0426$ & $0.0235$ \\
& $4$ & $3.62, 0.0607$ & $3.63$ & $0.0646$ & $0.0478$ & $0.0205$ \\ \hline

$\boldsymbol{c_{\mathrm{s}0} = 0.02}$, $\boldsymbol{\Gamma = 4/3}$
& $2$ & $1.78, 0.0408$ & $1.78$ & $0.0434$ & $-$ & $-$ \\
& $3$ & $2.69, 0.0512$ & $2.69$ & $0.0550$ & $-$ & $-$ \\
& $4$ & $3.62, 0.0591$ & $3.63$ & $0.0611$ & $-$ & $-$ \\ \hline

$\boldsymbol{c_{\mathrm{s}0} = 0.05}$, $\boldsymbol{\Gamma = 1}$
& $2$ & $1.62, 0.0588$ & $1.62$ & $0.0600$ & $0.0526$ & $0.0211$ \\
& $3$ & $2.48, 0.0716$ & $2.47$ & $0.0716$ & $0.0642$ & $0.0376$ \\
& $4$ & $3.36, 0.0800$ & $3.36$ & $0.0776$ & $0.0704$ & $0.0479$ \\ \hline

$\boldsymbol{c_{\mathrm{s}0} = 0.05}$, $\boldsymbol{\Gamma = 4/3}$
& $2$ & $1.62, 0.0553$ & $1.61$ & $0.0552$ & $-$ & $-$ \\
& $3$ & $2.48, 0.0682$ & $2.47$ & $0.0665$ & $-$ & $-$ \\
& $4$ & $3.36, 0.0763$ & $3.36$ & $0.0724$ & $-$ & $-$ \\ \hline

\end{tabular}
\end{center}
\caption{Summary of trapped $p$-mode simulations. From left to right, the columns give the equation of state ($c_{\mathrm{s}0}$ and $\Gamma$), azimuthal mode number $m$, the semi-analytic mode frequency $\omega_\mathrm{r}$ and growth rate $\omega_\mathrm{i}$ (zero viscosity), the numerically-calculated mode frequency, and the numerically-calculated growth rate for $\alpha = 0$ and several non-zero values of $\alpha$. The blank entries correspond to runs for which $p$-modes were not observed due to a second type of instability, as described in Section \ref{subsec:newtonian}.}
\label{tab:pmode_summary}
\end{table*}

\begin{figure}
\begin{center}
\includegraphics[width=0.49\textwidth]{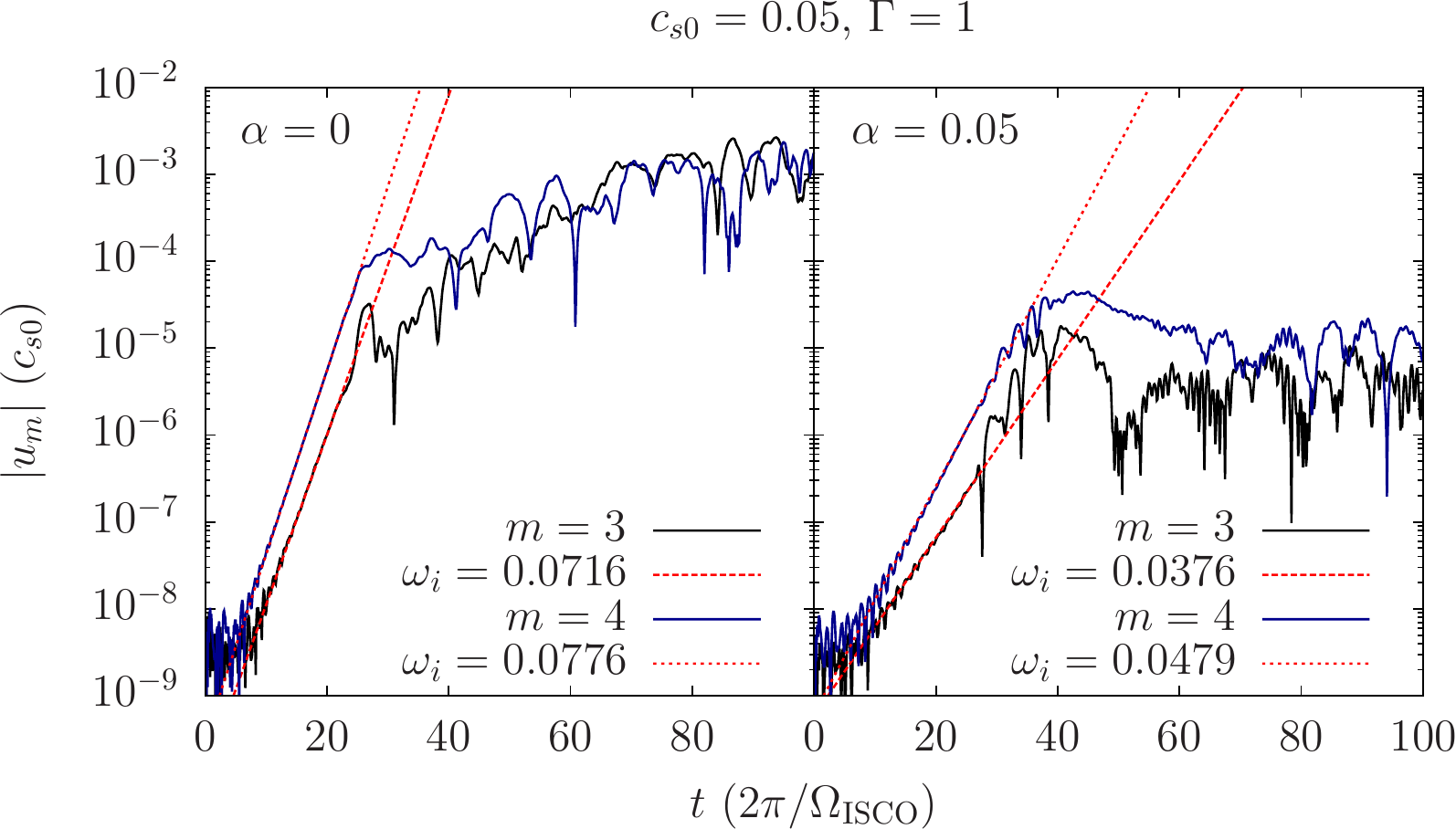}
\caption{Evolution of the $m = 3$ and $m = 4$ components of radial velocity [$|u_m|$, in units of $c_{\mathrm{s}0} = c_\mathrm{s}\left(r_\mathrm{ISCO}\right)$] at $r = 1.05$ for a typical run, showing exponential growth of trapped $p$-mode amplitudes, for zero and non-zero $\alpha$. The dashed and dotted lines show the exponential curves used to fit the mode growth rates. The reduction in growth rate and reduced saturation amplitude due to viscosity can be seen.}
\label{fig:growth_pmode}
\end{center}
\end{figure}

\begin{figure}
\begin{center}
\includegraphics[width=0.49\textwidth]{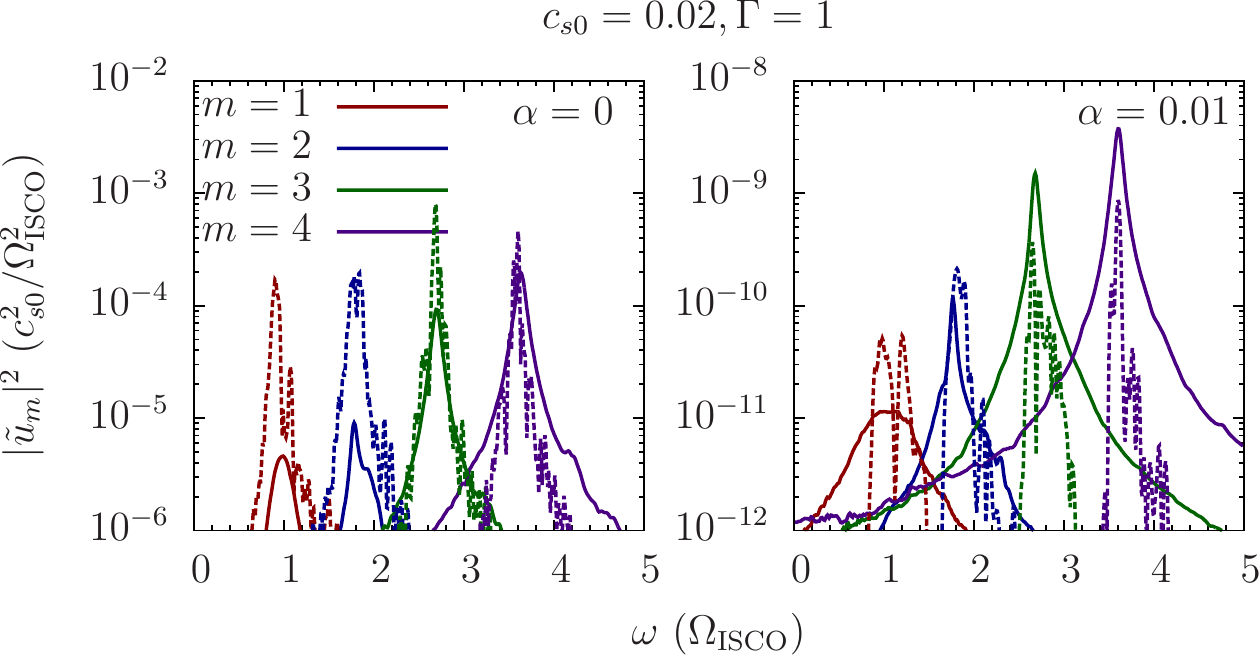}
\caption{Power spectra delineated by azimuthal number $m$ in the linear phase (solid curves) and nonlinear phase (dashed curves) for trapped $p$-modes with zero (left) and non-zero (right) $\alpha$. The linear phase power spectra have been multiplied by $10^{10}$ and $10^4$ (left and right, respectively) in order to compare them to the nonlinear phase power spectra.}
\label{fig:mfreq_pmode}
\end{center}
\end{figure}

\begin{figure}
\begin{center}
\includegraphics[width=0.49\textwidth]{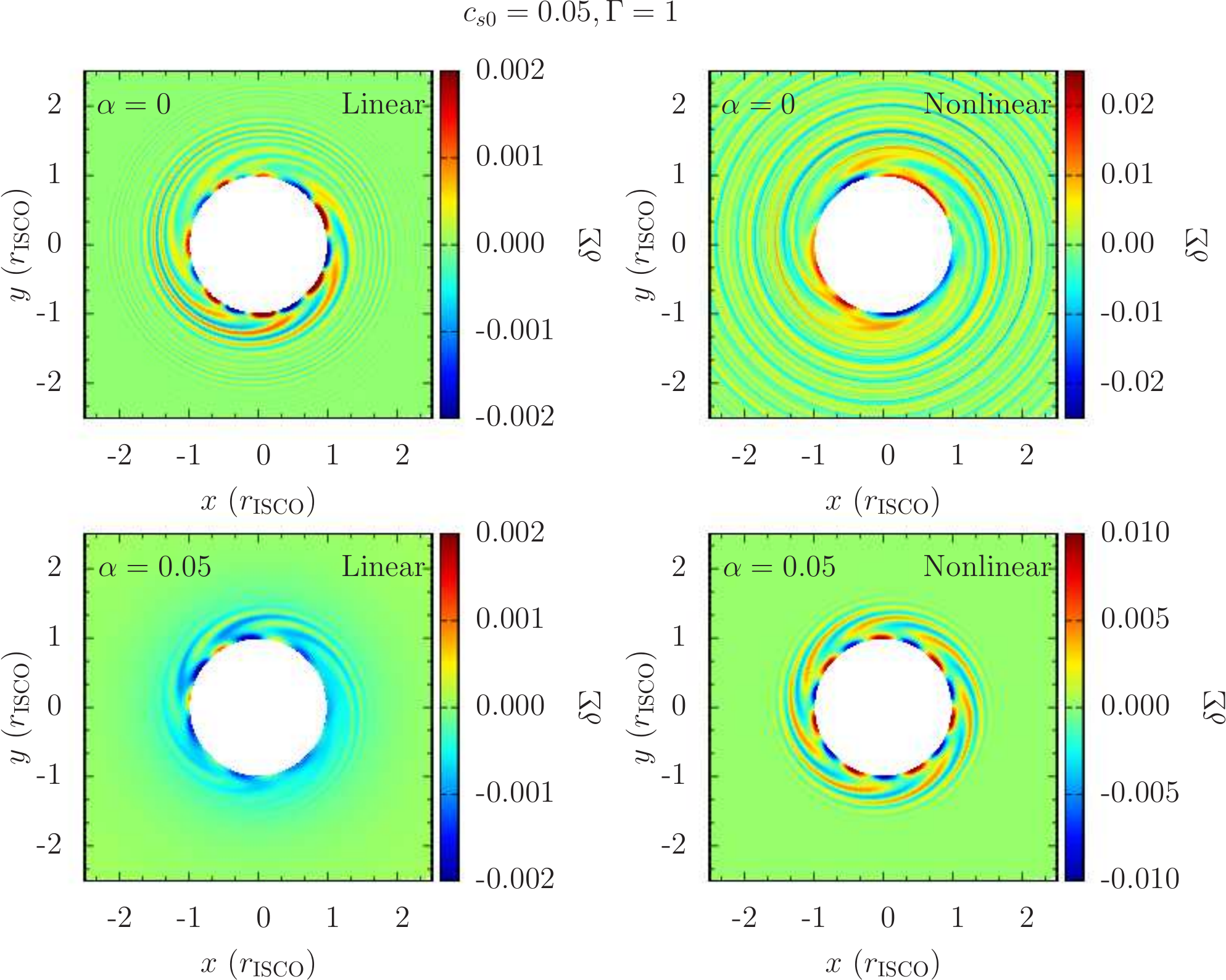}
\caption{Surface density perturbations near the end of the linear phase (left) and in the nonlinear phase (right) with zero (top) and non-zero (bottom) $\alpha$, showing the qualitative difference in saturation of trapped $p$-modes between the two cases.}
\label{fig:snapshot_pmode}
\end{center}
\end{figure}

We now investigate the behavior of trapped $p$-modes in a disc truncated at its inner edge by an impermeable wall. As noted before, such an inner disc edge mimics a magnetospheric boundary. We place the inner boundary at $r_\mathrm{ISCO}$ and impose a reflecting boundary condition with $u_r = 0$ (imposed by anti-symmetrizing $u_r$ across the boundary), and with all other variables fixed at their equilibrium values. In general, viscous forces induce a radial drift, which would lead to accumulation of mass at the inner boundary. We avoid this by initializing the surface density with a profile such that there is no radial drift (equation \ref{eq:radial_drift}), given by
\be
\Sigma \left(r\right) \propto \left[r^{-2}\left(\frac{r-r_\mathrm{s}}{3r-r_\mathrm{s}}\right)\right]^{1/\Gamma}.
\ee
This ensures that the azimuthally-averaged surface density does not change significantly as long as the perturbations to this equilibrium remain small (in principle this may be violated when the perturbation amplitudes become large and the modes are nonlinear). This initial surface density is modified by small ($\delta \Sigma / \Sigma \lesssim 10^{-6}$) random perturbations which serve as seeds for the growth of $p$-modes. They lead to the growth of many modes with many different values of $m$, the frequencies and growth rates of which can be measured independently from a single run (the accuracy of this approach is examined in Section \ref{subsec:pmode_resolution}).

Each mode is assumed to have a complex frequency $\omega = \omega_\mathrm{r} + i\omega_\mathrm{i}$ where $\omega_\mathrm{r}$ and $\omega_\mathrm{i}$ (both real) are the mode frequency and growth rate. The amplitude of radial velocity perturbations with azimuthal number $m$ at radius $r$ are determined numerically by computing
\be
u_m\left(r,t\right) = \frac{1}{2\pi}\int_{0}^{2\pi} u_r\left(r,\phi,t\right) e^{-im \phi} \mathrm{d}\phi.
\ee
By fitting a straight line to $\log|u_m|$ in the linear growth phase, we can determine the growth rate of each mode. The frequencies are determined using a power spectrum delineated by azimuthal number $m$ (equation \ref{eq:mfreq}). Some examples of analysis using these tools will be shown in the subsequent discussion of the results of the simulations.

Figure \ref{fig:growth_pmode} shows the evolution of $|u_m|$ (for $m = 3$ and $m = 4$, evaluated at a representative radius) for a pair of typical runs. Initially there is a linear growth phase lasting $20 - 40$ orbits, in which the amplitudes grow exponentially. The fitted curves used to measure the growth rates are shown. After the departure of the mode amplitudes from simple exponential growth, we refer to the rest of the evolution as the nonlinear phase. Example power spectra are shown in Figure \ref{fig:mfreq_pmode}, showing the mode frequencies in both the linear and nonlinear phases.

\subsection{Inviscid Case}
\label{subsec:pmode_inviscid}
We first perform simulations of an inviscid disc ($\alpha = 0$). Since the complex frequencies of the trapped $p$-modes can be compared to linear theory (Lai \& Tsang 2009), these serve as test cases for verifying the robustness and accuracy of the simulations before we investigate the effects of viscosity (see Fu \& Lai 2013 for a more detailed analysis of numerical simulations of trapped $p$-modes in inviscid discs). We choose four equations of state, with $c_{\mathrm{s}0} = 0.02$ or $c_{\mathrm{s}0} = 0.05$ and $\Gamma = 1$ or $\Gamma = 4/3$. The main results, numerically measured frequencies and growth rates of the modes, are presented in Table \ref{tab:pmode_summary}. For runs with $c_{\mathrm{s}0} = 0.02$, the numerical growth rates agree with those computed from linear theory to better than $10\%$, while for $c_{\mathrm{s}0} = 0.05$ they agree to better than $5\%$, with some cases in even better agreement. In all cases the numerical mode frequencies agree with linear theory to better than $1\%$, and these frequencies differ very little between the linear and nonlinear phases (see Figure \ref{fig:mfreq_pmode}). See Section \ref{subsec:pmode_resolution} for a resolution test demonstrating that these results are converged. Having accurately captured the properties of the inviscid trapped $p$-modes, we move on to exploring the effects of viscosity in the next subsection.

\subsection{Viscous Effects}
\label{subsec:pmode_viscous}
We carry out simulations using the same parameters described above but now with non-zero viscosity, choosing $\alpha = 0.01$ and $\alpha = 0.05$. Our analysis will focus on the modification of mode growth and saturation properties due to viscous effects. We separate this discussion into two parts based on the value of the polytropic index, as the effects of viscosity are different for $\Gamma = 1$ and $\Gamma = 4/3$.

The effect of viscosity on the mode growth rates are summarized in Table \ref{tab:pmode_summary}. We find qualitatively different results depending on the value of the polytropic index $\Gamma$. For $\Gamma = 1$, the growth rates of all modes are reduced compared to the inviscid case. For $c_{\mathrm{s}0} = 0.05$, the difference in growth rate due to viscosity $\delta \omega_\nu = \omega_\mathrm{i} - \omega_\mathrm{i}^{(\alpha=0)}$, is approximately proportional to $\alpha$, as expected theoretically (equation \ref{eq:Kato-formula}). This is not the case for $c_{\mathrm{s}0} = 0.02$, for which such an extrapolation of $\omega_\mathrm{i}$ based on the $\alpha = 0.01$ case would imply a negative value of $\omega_\mathrm{i}$ for $\alpha = 0.05$ (and thus the modes would not be overstable), which is not what we find. More importantly, the linear analysis of viscous effects predicts enhanced growth rates, rather than the diminished ones measured numerically. This indicates that the estimate of the viscous effect on the growth of global modes (equation \ref{eq:global-growthrates}) is inaccurate, perhaps due to dynamics of the corotation region which are neglected in equation (\ref{eq:global-growthrates}).

The saturation behavior of the the modes in the nonlinear phase is also affected by viscosity. Figure \ref{fig:growth_pmode} shows that while in the inviscid case, the various $u_m$ continue to grow slowly after the linear phase before saturating at larger amplitudes, with viscosity they decrease or level off after the linear phase, remaining somewhat steady at a relatively small amplitude. This effect can also be seen in Figure \ref{fig:snapshot_pmode}, which shows that when viscosity is included, the disc perturbations remain remarkably similar to the linear phase in appearance during the nonlinear phase, unlike the inviscid case for which the nonlinear phase looks qualitatively different than the linear phase. Therefore it appears that viscous forces ultimately suppress the mode amplitudes enough to prevent any significant nonlinear effects from occurring. 

Our simulations with reflective inner boundary yield unexpected results for $\Gamma = 4/3$. Unlike the case of $\Gamma = 1$ for which the overstable $p$-modes behave similarly with non-zero $\alpha$, albeit with reduced growth rates, when $\Gamma = 4/3$ we find no trace of the trapped $p$-modes. Instead, the power spectrum for these runs shows oscillations with frequencies approximately equal to $m\Omega_\mathrm{ISCO}$. For these runs, the evolution of $|u_m|$ does not show clean exponential growth as the $p$-modes do, but rather a sharp initial rise to a significant amplitude followed by slow incoherent growth. The amplitude of these oscillations eventually becomes unphysically large with radial velocities much greater than $c_{\mathrm{s}0}$. As the behavior of these unexpected modes is drastically different from the original $p$-modes, we forego detailed analysis of these runs. This highly unstable unexpected behavior may be the result of the combined effects of overstability due to GR (resulting from the smallness of $\kappa/\Omega$ and increased $q$) and due to viscosity (resulting from large $A$), both of which enhance the growth rate as given in eq. (\ref{eq:Kato-formula}). Since we are interested in the viscous effect, in Section \ref{subsec:newtonian} we will isolate it by modifying the numerical setup to simulate a disc with a pure Newtonian $1/r$ potential, to see if this behavior persists in a simple way that allows for a concrete analysis.

\subsection{Viscosity-Driven Boundary Modes in Newtonian Potential}
\label{subsec:newtonian}

\begin{figure}
\begin{center}
\includegraphics[width=0.49\textwidth]{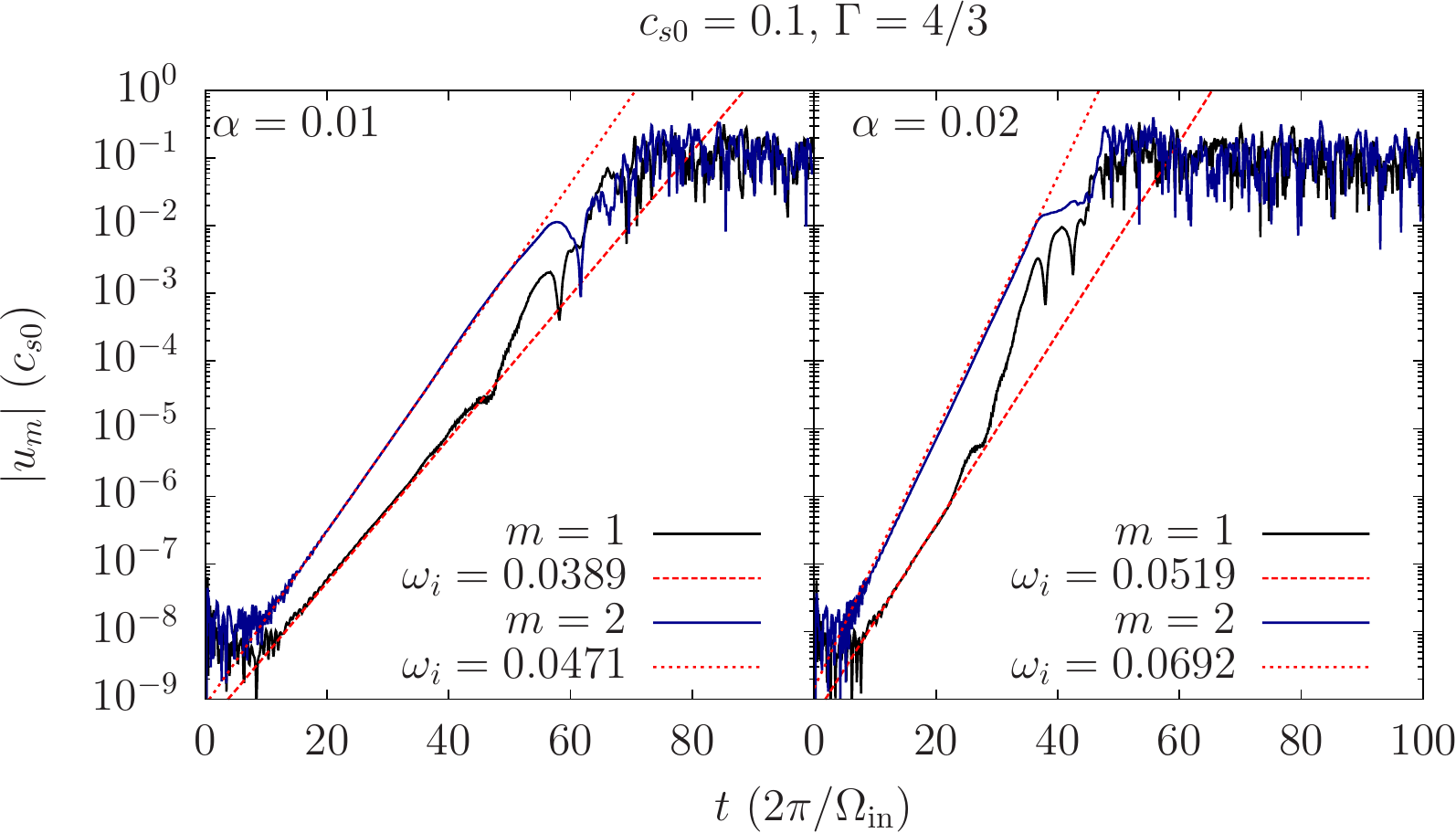}
\caption{Evolution of $|u_m|$ at $r = 1.05$ for Newtonian disc with reflective inner boundary. The $m = 1$ and $m = 2$ components are shown, along with exponential fits used to determine growth rates, for two values of $\alpha$. The growth rates increase with increasing $\alpha$, in qualitative agreement with the theory of viscous overstability.}
\label{fig:growth_newton}
\end{center}
\end{figure}

\begin{figure}
\begin{center}
\includegraphics[width=0.49\textwidth,clip]{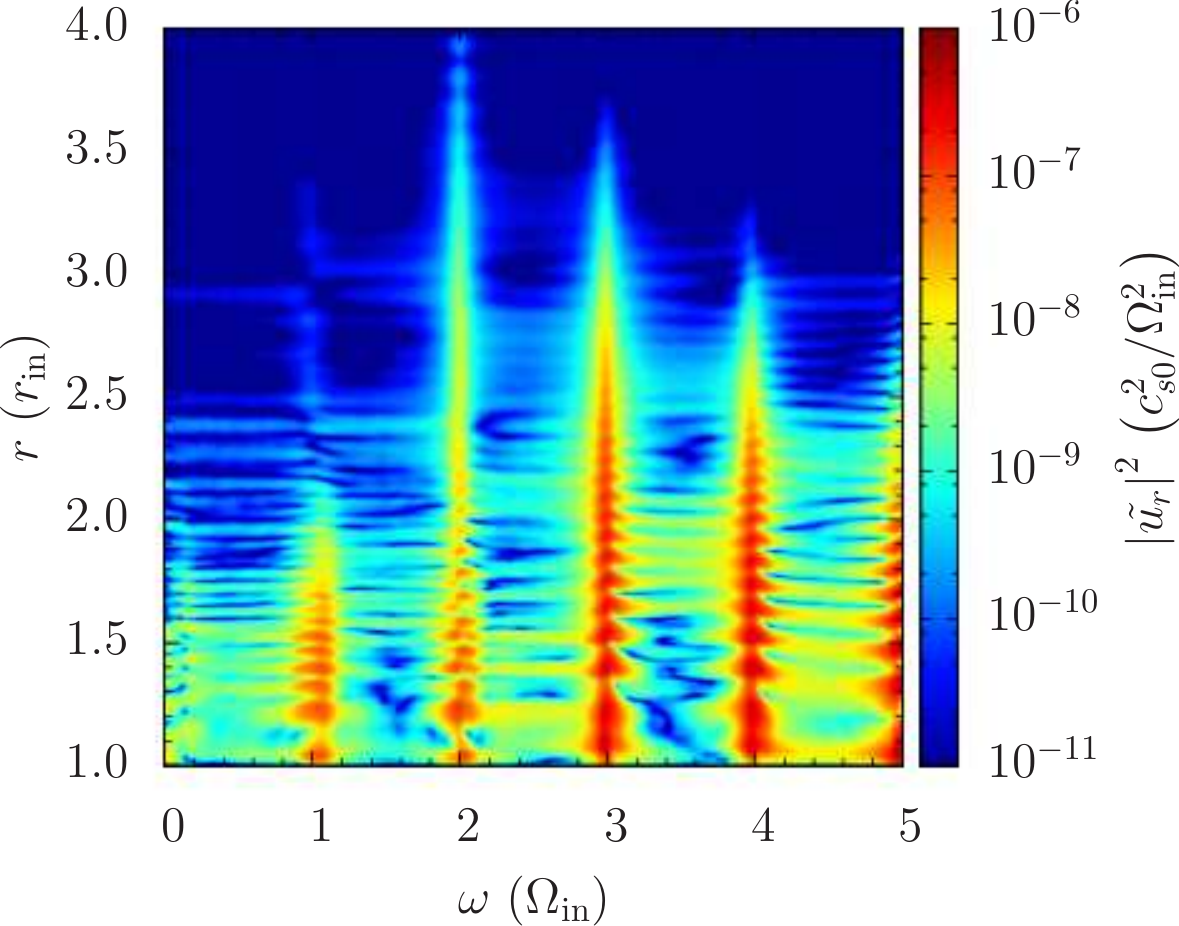}
\caption{Linear phase power spectrum for viscosity-driven boundary modes in Newtonian disc.}
\label{fig:psd_newton}
\end{center}
\end{figure}

\begin{figure}
\begin{center}
\includegraphics[width=0.49\textwidth]{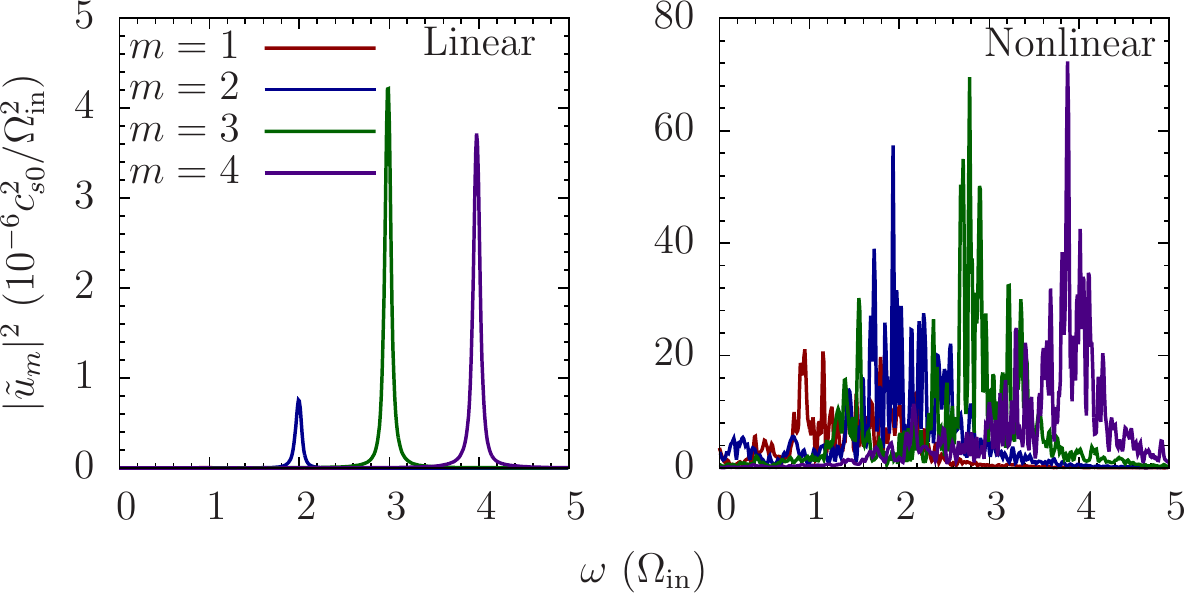}
\caption{Power spectra for lowest $m$ boundary modes in linear and nonlinear phases. In the linear phase, each mode corresponds to a single sharp frequency peak close to $m \Omega_\mathrm{ISCO}$, while in the nonlinear phase these peaks are broader and exhibit splittings, but retain powers concentrated in frequencies similar to those found in the linear phase.}
\label{fig:mfreq_newton}
\end{center}
\end{figure}

\begin{figure}
\begin{center}
\includegraphics[width=0.49\textwidth,clip]{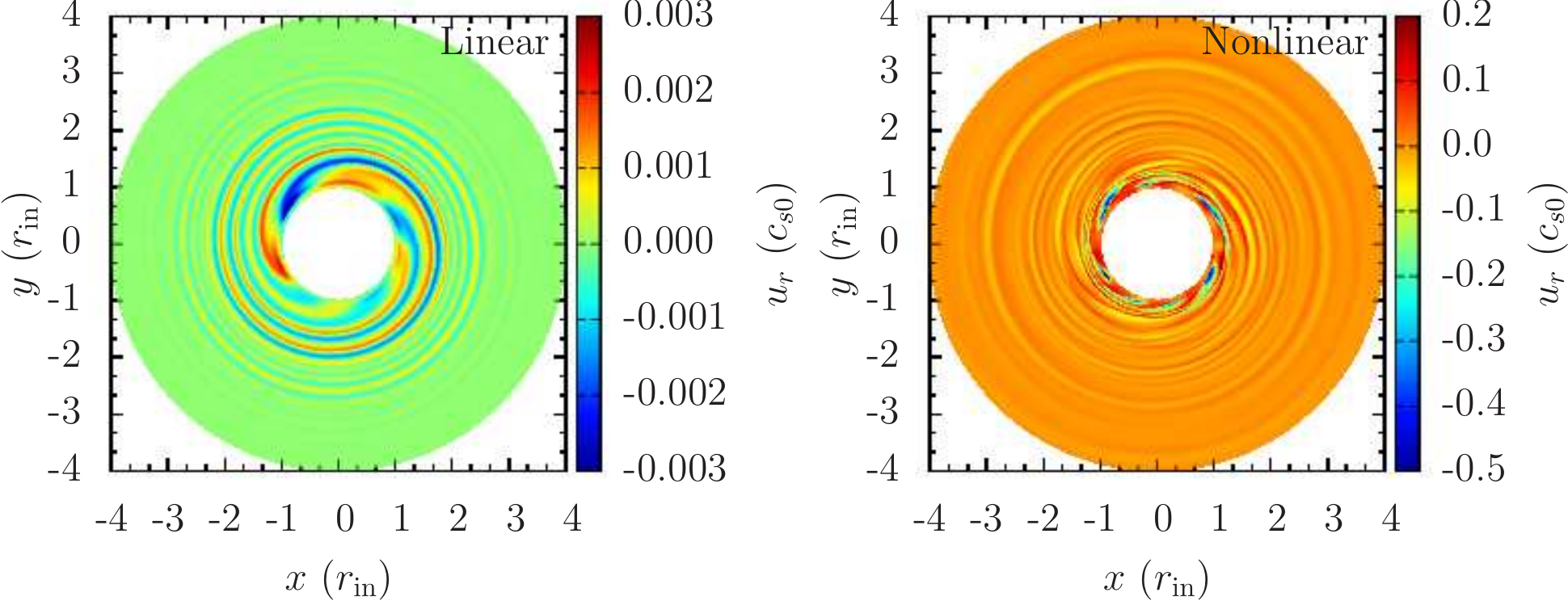}
\caption{Radial velocity of boundary modes in Newtonian discs in the linear and nonlinear phase.}
\label{fig:snapshot_newton}
\end{center}
\end{figure}

\begin{table}
\begin{center}
\begin{tabular}{|c|c|c|c|}
\hline
\hline
& $m$   & $\omega_\mathrm{r}$ & $\omega_\mathrm{i}$ \\ \hline
$\alpha = 0.01$ & $1$ & $1.05$ & $0.0389$ \\
                & $2$ & $2.00$ & $0.0471$ \\
                & $3$ & $2.99$ & $0.0488$ \\
                & $4$ & $3.98$ & $0.0488$ \\
$\alpha = 0.02$ & $1$ & $1.01$ & $0.0519$ \\
                & $2$ & $2.02$ & $0.0692$ \\
                & $3$ & $3.01$ & $0.0743$ \\
                & $3$ & $4.00$ & $0.0744$ \\ \hline
\end{tabular}
\end{center}
\caption{Frequencies and growth rates of boundary modes.}
\label{tab:newton_summary}
\end{table}

We wish to determine whether or not the growth of the new modes which overwhelm the $p$-modes is due to a purely viscous effect which can operate in the absence of GR effects. To this end, we perform simulations using the same setup as described at the beginning of this section, but replace the pseudo-Newtonian gravitational potential with a pure Newtonian potential, $\Phi = -GM/r$. Under this potential there is no longer an ISCO, so the location of the inner boundary no longer has any physical significance, and we simply denote it as $r_\mathrm{in}$. The corresponding boundary condition, as in the beginning of this section, is reflective with $u_r = 0$. We choose four sets of parameters, using $c_{\mathrm{s}0} = 0.1$ and $\left(\Gamma,\alpha\right) = \left(1,0\right), \left(1,0.01\right), \left(4/3,0\right), \left(4/3,0.01\right)$. Due to the larger sound speed used, and therefore larger radial wavelengths, we reduce our resolution to $N_r \times N_\phi = 512 \times 128$. 

We observe no growing oscillations for any of these runs except for $\left(\Gamma,\alpha\right) = \left(4/3,0.01\right)$, confirming that the anomalous modes observing in Section \ref{subsec:pmode_viscous} can be attributed purely to viscous effects (that is, requires $\alpha > 0$ and $\Gamma > 1$) and do not depend on GR effects. We therefore performed an additional run with $\left(\Gamma,\alpha\right) = \left(4/3,0.02\right)$ to probe their dependence on $\alpha$. Figure \ref{fig:growth_newton} shows the evolution of the lowest $m$ mode amplitudes (represented by $|u_m|$ at $r = 1.05$), showing their exponential growth at a rate which increases with $\alpha$ (see Table \ref{tab:newton_summary} for complete list of frequencies and growth rates), as well as their relatively flat saturation amplitude consistent with the the viscous effect on mode saturation seen in Section \ref{subsec:pmode_viscous}. Figure \ref{fig:psd_newton} shows the power spectrum for these modes, indicating that they are global, with power in discrete frequencies across a range of radii. This justifies the use of eq. (\ref{eq:mfreq}) to produce a radially integrated $m$-delineated power spectrum, as shown in Figure \ref{fig:mfreq_newton}. From this we see that in the nonlinear phase, the modes retain approximately the same frequencies as in the linear phase, but with broader power spectra peaks and the presence of other sub-dominant peaks at new frequencies. The morphology of the two phases is shown in Figure \ref{fig:snapshot_newton}. Note that the linear phase bears some resemblance to that of the $p$-modes (compare to Figure \ref{fig:snapshot_pmode}), while the nonlinear phase appears very different, having a more complex appearance.

These modes may be related to the interface modes studied by Tsang \& Lai (2009b) and Fu \& Lai (2012), who considered a Keplerian disc in contact with a uniformly rotating cylinder (for example, a magnetosphere), which is similar to the setup we consider. Their results indicate that modes with frequencies close to $m\Omega_\mathrm{in}$ can be overstable due to the corotation resonance effect. In our case, they may not be intrinsically overstable but are only driven by viscous forces which can act in phase with the oscillations, by having a sufficiently large $A$ parameter. A qualitative description of this phenomenon is as follows. In our simulation, the disc is initialized with a surface density for which viscous forces induce zero radial drift. However, a small perturbation to this equilibrium leads to a non-zero $u_r$, which leads to a radial oscillation at the local epicyclic frequency (which is simply equal to the orbital frequency $\Omega$ in a Newtonian potential). When material close to the inner solid boundary is pushed inward due to this oscillation, it is compressed, leading to a stronger viscous forces, which act as an additional restoring force (in addition to gravity which is producing the epicyclic motion). In this way, the restoring force increases with the amplitude of the oscillation, leading to overstability.

\subsection{Resolution Study}
\label{subsec:pmode_resolution}

\begin{table}
\begin{center}
\begin{tabular}{|c|c|c|c|}
\hline
\hline

$\alpha$ & $\left(\omega_\mathrm{r},\omega_\mathrm{i}\right)_\mathrm{half}$ & $\left(\omega_\mathrm{r},\omega_\mathrm{i}\right)_\mathrm{standard}$ & $\left(\omega_\mathrm{r},\omega_\mathrm{i}\right)_\mathrm{double}$ \\ \hline

$0$ & $1.79, 0.0396$ & $1.78, 0.0419$ & $1.78, 0.0423$ \\

$0.01$ & $1.77, 0.0301$ & $1.76, 0.0315$ & $1.76, 0.0317$ \\ \hline

\end{tabular}
\end{center}
\caption{Frequency and growth rate of $m = 2$ trapped $p$-mode for $\left(c_{\mathrm{s}0},\Gamma\right) = \left(0.02,1\right)$ with initial $m = 2$ perturbation for half, standard and double resolution runs.}
\label{tab:pmode_resolution}
\end{table}

As in Section \ref{subsec:ts_resolution}, we perform two resolution test runs with half ($512 \times 128$) and double ($2048 \times 512$) the standard resolution, for the case $\left(c_{\mathrm{s}0},\Gamma,\alpha\right) = \left(0.02,1,0\right)$. These runs also differ from the standard run in that the initial density perturbations are proportional to $\cos\left(2 \phi\right)$, so that the $m = 2$ mode is isolated. This allows us to test the accuracy of measuring mode frequencies and growth rates for modes of multiple $m$ from an initially random perturbation, as is done in our standard runs.  The results are shown in Table \ref{tab:pmode_resolution}. We see that with increasing resolution, the numerically measured growth rate converges on the result from linear theory, agreeing to better than $1\%$ for the double resolution run. Additionally, we see that a standard resolution run with a pure $m = 2$ initial perturbation reproduces the growth rate of this mode more accurately than a run with random perturbations, as used in Section \ref{subsec:pmode_viscous}. However, the difference is small enough to justify the use of the random perturbation approach, which allows us to capture multiple modes from single runs. We note that since this resolution test was performed for our smallest sound speed, for which the modes have the smallest radial wavelengths, these tests provide a lower limit on our accuracy, as we expect the longer wavelength modes associated with our larger sound speed to be better resolved by the same resolution. We also perform the same test using $\left(c_{\mathrm{s}0},\Gamma,\alpha\right) = \left(0.02,1,0\right)$ to ensure that viscous effects are not sensitive to resolution. For this case we achieve similarly excellent convergence of growth rate with increasing resolution. Additionally, the discrepancy between the run with random perturbations and the one with $m = 2$ perturbations is smaller than the inviscid case.

\section{Discussion}
\label{sec:discussion}

We have carried out viscous hydrodynamical simulations of black hole
(BH) accretion discs to determine how viscosity may drive global
oscillation modes in the disc. Analytical considerations (see
Section \ref{sec:theory}) indicate that, depending on the viscosity law (particularly
how viscosity scales with density) and the background disc flow
properties, various global oscillations with different azimuthal
wavenumbers may be excited. We considered two types of accretion
flows. The first involves discs with transonic radial inflow across the
innermost stable circular orbit (ISCO) of the BH, and the second involves
discs truncated at an inner edge and with a free outer boundary 
at large distance.

Our simulations on transonic discs (Section \ref{sec:transonic}) follow previous works in one
dimension (Milsom \& Taam 1996) and in two or three
dimensions (O'Neill et al.~2009; Chan 2009). We show that, depending
on the viscosity parameter $\alpha$, global oscillations with
different azimuthal wavenumbers $m$, and with frequencies close to
$\kappa_{\rm max}$, $m\Omega_{\rm ISCO}$ or their linear combinations
(where $\kappa_{\rm max}$ and $\Omega_{\rm ISCO}$ are the maximum
radial epicyclic frequency and the orbital frequency at the ISCO,
respectively) are excited. In general, as $\alpha$ increases above a critical value
$\alpha_{\rm crit 1}$ ($\sim 0.1 - 0.25$ for isothermal discs),
non-axisymmetric modes (typically $m=4$) first develop, 
then axisymmetric modes develop above a second critical value $\alpha_{\rm crit 2}$ ($\sim 0.5$).
We propose that these modes are the result of viscous overstability, possibly
combined with an effect related to instability of the sonic point, although
the theory of the latter is not strictly applicable in our simulations.

Our simulations on truncated discs (Section \ref{sec:trapped}) extend previous semi-analytic 
works on overstable inertial-acoustic modes ($p$-modes) driven by 
corotation resonance in non-magnetic discs (Lai \& Tsang 2009;
Tsang \& Lai 2008, 2009b; Horak \& Lai 2013)
and magnetic discs (Fu \& Lai 2009; 2011), on magnetosphere-disc interface
modes (Tsang \& Lai 2009a; Fu \& Lai 2012), and numerical studies of
nonlinear $p$-modes (Fu \& Lai 2013).  We find that the growth rates of overstable $p$-modes
are reduced by viscosity in isothermal discs, while for other equations of state 
(and corresponding viscosity laws), they are suppressed
due to the excitation of a different class of modes with frequencies $\omega \approx m \Omega_\mathrm{in}$, which are related to magnetosphere-disc interface modes.

Obviously, our 2D hydrodynamical disc models do not capture various
complexities (e.g., magnetic field, turbulence and radiation)
associated with real BH accretion discs.  Much progress in numerical
General Relativistic Magnetohydrodynamic (GRMHD) simulations of BH
accretion discs has been made in the past decade, but much work
remains to understand the complex phenomenology of BH X-ray
binaries (e.g., Remillard \& McClintock 2006; Done et al.~2007; 
Belloni et al.~2012). Several recent simulations have revealed quasi-periodic
variabilities of various fluid variables, but the connection of these
variabilities to the observed HFQPOs is far from clear (e.g., Henisey
et al. 2009; O’Neill et al. 2011; Dolence et al. 2012; McKinney et
al. 2012; Shcherbakov \& McKinney 2013). Our simulations, although based on highly simplified disc models, 
demonstrate that under appropriate conditions, various global oscillation
modes can grow to nonlinear amplitudes with well-defined frequencies. We emphasize that our height-integrated treatment
cannot capture modes with dependence on vertical structure ($g$-modes and $c$-modes), which may be present in 3D simulations. The properties and excitation of the $p$-modes considered
in this paper may also be modified in a fully 3D approach.

It is of interest to compare the frequencies of our simulated modes to those of observed HFQPOs in X-ray binaries (Remillard \& McClintock 2006; Belloni et al. 2012), particulary the three systems whose BH spins are constrained using the continuum-fitting method (see Narayan \& McClintock 2012 and references therein). These considerations indicate that the frequencies of the HFQPOs are smaller than the ISCO frequency, therefore they are unlikely to be explained by any non-axisymmetric modes with $\omega \gtrsim \Omega_\mathrm{ISCO}$. However, this does not rule out the axisymmetric $\kappa_\mathrm{max}$ modes which have lower frequencies. We note that we have not addressed the mechanism by which disc oscillations may manifest as variabilities in X-ray flux, which affects the observability (or lack of observability) of any oscillations that occur in the disc.

Although our simulations cannot be directly compared with
observations, some tentative conclusions can be drawn.  The thermal
(high-soft) state of BH X-ray binaries may be approximately described
by the transonic disc model (e.g. Done et al.~2007).  Our simulations
show that such a disc does not allow trapping of $p$-modes, but can
excite global $\kappa_{\rm max}$-waves or $m\Omega_{\rm ISCO}$-waves
when the viscosity parameter $\alpha$ is sufficiently large.  Since no
significant HFQPOs are observed in the thermal state of X-ray binaries,
our simulations provide an upper limit on the effective disc 
viscosity parameter of $\alpha \lesssim 0.5$.

The physical nature of the intermediate state (or ``steep power law
state'') of BH X-ray binaries, when HFQPOs are observed, is currently
uncertain (Done et al.~2007; Oda et al.~2010). Since episodic jets are
produced in this state, large-scale magnetic fields likely play an
important role (e.g., Tagger \& Varniere 2006; Yuan et al.~2009;
McKinney et al.~2012). When magnetic fields advect inwards in the
accretion disc and accumulate around the BH (e.g. Bisnovatyi-Kogan \&
Ruzmaikin 1974, 1976; Igumenshchev et al. 2003; Rothstein \& Lovelace
2008), a magnetosphere may form. The truncated disc model studied in
this paper may mimic the disc outside the magnetosphere. Our
calculations show that overstable $p$-modes (driven by corotation
resonance) can be trapped in the inner disc, but their growth rates
are typically reduced by disc viscosity. More importantly, with
increasing $\alpha$ and more general equation of state, a new series
of global waves can be excited at the disc boundary (see Tsang \& Lai
2009a; Fu \& Lai 2012; McKinney et al.~2012).

\section*{Acknowledgments}

We thank David Tsang for useful discussion.  This work has been
supported in part by NSF grants AST-1008245, 1211061, and NASA grant
NNX12AF85G. J.H. acknowledges support from grants LH 14049 and M100031202.

\appendix

\section{Wave Equation Up to First Order in Viscosity and Radial Infall Velocity}
\label{sec:Appendix1}
In this appendix we summarize the procedure of deriving the equations (\ref{eq:visc-wave-eq})--(\ref{eq:L-v}). The linearly perturbed continuity equation (\ref{eq:continuity-0}) and the Navier-Stokes equation (\ref{eq:Navier-Stokes-0}) in the polar coordinates $(r,\phi)$ read
\be
	-\ii\tilde{\omega}\delta\Sigma + \frac{1}{r}\partial_r\left(r\Sigma\delta u_r\right) + 
	\frac{\ii m}{r}\Sigma\delta u_\phi 
	= -\frac{1}{r}\partial_r\left(r u^r \delta\Sigma \right) \equiv \delta\dot{\Sigma}
	\label{eq:pert-rho}
\ee
\be
      -\ii\tilde{\omega}\delta u_r - 2\Omega\delta u_\phi + \delta h_{,r} = 
	-\partial_r\left(u_r\delta u_r\right) + 
	\delta\left(\frac{1}{\Sigma}\mathcal{F}_r^\mathrm{visc}\right)
	\equiv\delta\dot{u}_r,
	\label{eq:pert-r}
\ee
\be
	-\ii\tilde{\omega}\delta u_\phi + \frac{\kappa^2}{2\Omega}\delta u_r + 
	\frac{\ii m}{r}\delta h =
	-\frac{u_r}{r}\partial_r\left(r\delta u_\phi\right) + 
	\delta\left(\frac{1}{\Sigma}\mathcal{F}_\phi^\mathrm{visc}\right)
	\equiv\delta\dot{u}_\phi,
	\label{eq:pert-phi}
\ee
The equations were arranged in such a way that the coefficients on the LHS do not depend on the viscosity, while the RHS terms, denoted by $\delta\dot{\Sigma}$, $\delta\dot{u}_r$ and $\delta\dot{u}_\phi$, are at least linear in the viscosity $\eta$. The first RHS term in each equation describes the effect of the radial inflow in the stationary case, the second terms in the last two equations are contributions due to the perturbation of the viscous force $\vec{\mathcal{F}}^\mathrm{visc}=\vec{\nabla}\cdot\boldsymbol{\sigma}$.

In what follows we would like to express the RHS terms $\delta\dot{u}_r$, $\delta\dot{u}_\phi$ and $\delta\dot{\Sigma}$ in terms of the velocity perturbation $\delta u^r$, $\delta u^\phi$ and enthalpy $\delta h$. To simplify the analysis we will further work in the leading order of the WKBJ approximation. Therefore, we assume that the radial wavelength $\lambda_r = 2\pi/k_r$ of the perturbations is much smaller than the radius $r$. More specifically, we assume that $k_r r = \mathcal{O}(r/H)$, where $H$ is the disc semi-thickness. In thin discs, this assumption is satisfied everywhere except for small regions close to the Lindblad and corotation resonances. In the leading order in the ratio $r/H$, we obtain
\begin{eqnarray}
	\delta\dot{\Sigma} &\approx& 
	-u_r\frac{\Sigma}{c_\mathrm{s}^2}\frac{\dd}{\dd r}\delta h_{,r}
	\label{eq:rhs-rho}
	\\
	\delta\dot{u}_r &\approx& 
	\left[-u_r \frac{\dd}{\dd r} + \frac{4}{3}\nu\frac{\dd^2}{\dd r^2}\right] \delta u_r,
	\label{eq:rhs-r}
	\\
	\delta\dot{u}_\phi &\approx& 
	\left[-u_r \frac{\dd}{\dd r} + \nu \frac{\dd^2}{\dd r}\right] \delta u_\phi -
	\nu q A \frac{\Omega}{c_\mathrm{s}^2} \frac{\dd}{\dd r} \delta h.
	\label{eq:rhs-phi}
\end{eqnarray}
In all three equations, the first terms are proportional to $u_r$ and describe the effects of the radial inflow. The second terms in equation (\ref{eq:rhs-r}) and (\ref{eq:rhs-phi}) are the contributions of the perturbed visous force due to velocity field connected to the wave. Finally, the third term of equation (\ref{eq:rhs-phi}) results from the change of the viscous force due to perturbation of the dynamic viscosity coefficient $\eta$. 

By substituting expressions (\ref{eq:rhs-r}) and (\ref{eq:rhs-phi}) into the equations (\ref{eq:pert-r}) and (\ref{eq:pert-phi}) and expanding the velocity perturbation $\delta\vec{u}$ in powers of $\nu$, 
\begin{equation}
	\delta u_r = \delta u_r^{0} + \nu\delta u_r^{1} + \dots,
	\quad
	\delta u_\phi = \delta u_\phi^{0} + \nu\delta u_\phi^{1} + \dots,
\end{equation}  
we may solve perturbativelly these equations for $\delta\vec{u}$ in terms of the enthalpy perturbation $\delta h$. Up to the first order in $\nu$ (and $u_r$), we obtain
\begin{eqnarray}
	\delta u_r &=& \frac{\ii}{D}\left(\tilde{\omega}\frac{\dd}{\dd r} - \frac{2m\Omega}{r}\right)\delta h
	-\frac{u_r}{D^2}\left(\kappa^2 + \tilde{\omega}^2\right)\frac{\dd^2}{\dd r^2}\delta h +
	\nonumber\\ &\phantom{=}& 
	\frac{\nu}{D^2}\left[\left(\kappa^2+\frac{4}{3}\tilde{\omega}^2\right)\frac{\dd^3}{\dd r^3}- 
	2qA\Omega^2\frac{D}{c_\mathrm{s}^2}\frac{\dd}{\dd r}\right]\delta h + \mathcal{O}(\nu^2), 
	\label{eq:du-r}
	\nonumber\\
	\delta u_\phi &=& \frac{1}{D}\left(\frac{\kappa^2}{2\Omega}\frac{\dd}{\dd r} - \frac{m\tilde{\omega}}{r}\right)\delta h +
	\ii u_r\frac{\kappa^2\tilde{\omega}}{2\Omega D^2}\frac{\dd^2}{\dd r^2}\delta h -
	\nonumber\\ &\phantom{=}& 
	\ii\tilde{\omega}\frac{\nu}{D^2}\left[
	\frac{7}{3}\frac{\kappa^2}{2\Omega}\frac{\dd^3}{\dd r^3} + 
	qA\Omega\frac{D}{c_\mathrm{s}^2}\frac{\dd}{\dd r}\right]\delta h + \mathcal{O}(\nu^2).
	\label{eq:du-phi}
\end{eqnarray}
Finally, by substituting these expressions into the perturbed continuity equation (\ref{eq:pert-rho}), we recover the desired equation (\ref{eq:visc-wave-eq}) with operators $\hat{L}^0$ and $\hat{L}_\mathrm{in}^1$ and $\hat{L}_\mathrm{v}^1$. given by equations (\ref{eq:inviscid-wave-eq}) and (\ref{eq:L-in}) and (\ref{eq:L-v}). We remind the reader that these equations are valid only in the leading order in both $(H/r)$- ratio, radial drift velocity $u_r$ and viscosity $\nu$ under the assumpion of short radial wavelength (as compared to the radial coordinate $r$). Hence, in principle, it cannot be applied in the regions of the Lindblad and corotation resonances. A proper description of those effects would require keeping also the singular terms in equations (\ref{eq:rhs-rho})--(\ref{eq:du-phi}).

\section{Growth Rates of Modes}
\label{sec:Appendix2}
We would like to find the \textit{global} change of the eigenfrequencies of $p$-modes trapped in the inner disc. In the following, we will show that it is given by some proper radial average of the Kato (1978) local growth rate over the propagation region. Introducing $\psi=S^{-1/2}\delta h$ with $S=D/(r\Sigma)$, equation (\ref{eq:visc-wave-eq}) takes the form
\begin{equation}
	\left[\frac{\dd^2}{\dd r^2} - V(r)\right] \psi + S^{-1/2}\hat{L}^1\left(S^{1/2} \psi\right) = 0,
\end{equation} 
where
\begin{equation}
	V(r) = \frac{D}{c_\mathrm{s}^2} + \frac{2m\Omega}{r\tilde{\omega}}\frac{d}{dr}\ln
	\left(\frac{\Omega\Sigma}{D}\right) + S^{1/2}\frac{d^2}{dr^2} S^{-1/2} + \frac{m^2}{r^2},
\end{equation}
and $\hat{L}^1$ is either $\hat{L}^1_\mathrm{v}$, or $\hat{L}^1_\mathrm{in}$ or their sum. Assuming that the region of interest does not contain the resonances, the potential $V(r)$ can be approximated by just the first term because the others are by factor of order $(H/r)^2$ smaller. If we express $\psi$ and $D$ as $\psi = \psi_0 + \psi_1$ and $D=D_0 + D_1$ (where $\psi_1$ and $D_1$ are of order $\nu$, note that $D_1 = -2\tilde{\omega}_0\omega_1$), we obtain in the zero order equation,
\begin{equation}
	\left[\frac{\dd^2}{\dd r^2} - \frac{D_0}{c_\mathrm{s}^2}\right] \psi_0 = 0,
	\label{eq:eigen-0}
\end{equation}
and the first order equation,
\begin{equation}
	\left[\frac{\dd^2}{\dd r^2} - \frac{D_0}{c_\mathrm{s}^2}\right] \psi_1 - 
	\frac{D_1}{c_\mathrm{s}^2} \psi_0 + S^{-1/2}\hat{L}^1\left(S^{1/2} \psi_0\right) = 0.
	\label{eq:eigen-1}
\end{equation} 
Multiplying by $\psi_0^\ast$ and integrating between radii $r_1$ and $r_2$ (trapping range of the mode), we obtain
\begin{equation}
\begin{aligned}
	&2\omega_1\int_{r_1}^{r_2}\frac{\tilde{\omega}_0}{c_\mathrm{s}^2}\left|\psi_0^2\right| \dd r + 
	\int_{r_1}^{r_2} \psi_0^\ast\left(S^{-1/2} \hat{L}^1 S^{1/2} \psi_0\right) \dd r \\ 
	&= -\int_{r_1}^{r_2} \psi_0^\ast \left[\frac{\dd^2}{\dd r^2} - 
	\frac{D_0}{c_\mathrm{s}^2}\right] \psi_1 dr \equiv
	\mathcal{I}.
\end{aligned}
\label{eq:integ-1}
\end{equation}
The integral on the RHS can be written as (after two integrations by parts)
\begin{equation}
	\left[\psi_1 \frac{\dd \psi_0^\ast}{\dd r}-
	\psi_0^\ast \frac{\dd \psi_1}{\dd r}\right]_{r_1}^{r_2}
	-\int_{r_1}^{r_2} \psi_1\left[\frac{\dd^2}{\dd r^2} - 
	\frac{D_0}{c_\mathrm{s}^2}\right]\psi_0^\ast \dd r.
\end{equation}
Assuming that $\omega_0$ is real, the first integral vanishes. The second term is a factor $\mathcal{O}(H/r)$ smaller than the other integrals in equation (\ref{eq:integ-1}). Then, neglecting $\mathcal{I}$, we obtain
\begin{equation}
\label{eq:omega_1}
	\omega_1 = -\frac{\int \psi_0^\ast \left(S^{-1/2} \hat{L}^1 S^{1/2} \psi_0\right) \dd r}{
	2\int \left(\tilde{\omega}_0/c_\mathrm{s}^2\right) |\psi_0|^2 \dd r}.
\end{equation}

An approximate WKBJ solution of equation (\ref{eq:eigen-0}), valid up to the second order, is
\begin{equation}
\begin{aligned}
& \psi_0 = \frac{1}{\sqrt{k}}\left[\mathcal{A}_{-}\exp\left(-\ii\int^r k dr\right) + \mathcal{A}_{+}\exp\left(\ii\int^r k dr\right)\right], \\
& k = \sqrt{-D}/c_\mathrm{s},
\end{aligned}
\end{equation}
for which $|\psi_0|^2 = (|\mathcal{A}_{-}|^2 + |\mathcal{A}_{-}|^2)/k$. Evaluating the two integrals in equation (\ref{eq:omega_1}) for this solution, we find that
\begin{equation}
  \omega_1 = \frac{\int\delta\omega(r) w(r) dr}{\int w(r) dr},
  \quad
  w(r) = \frac{1}{c_\mathrm{s}}\frac{\tilde{\omega}_0}{\sqrt{-D}}.
  \label{eq:growthrate-global}
\end{equation}
Hence, the global change of the frequency of the mode is just the averaged local changes $\delta\omega_\mathrm{v}(r)$ or $\delta\omega_\mathrm{in}(r)$ of equations (\ref{eq:Kato-formula}) and (\ref{eq:inflow-formula}) with the weight function $w(r)$. 

Although expression (\ref{eq:growthrate-global}) does not apply close to the Lindblad resonances (LRs), we believe that the first-order LRs can be included as well, because the surrounding regions give negligible contributions to both integrals [$w(r)$ is integrable close to the first-order LRs]. Therefore, the integrals can be extended to the propagation regions of the oscillations. The lower integral however diverges for the second-order LR (at maxima of $\Omega + \kappa/m$) and more sophisticated analysis is needed in those cases.


\begin{thebibliography}{}
\bibitem[]{} Abramowicz, M. A., Klu\'{z}niak, W., 2001, A\&A, 374, L19
\bibitem[]{} Abramowicz, M. A., Klu\'{z}niak, W., Bursa, M., Hor\'{a}k, J., Rebusco, P., T\"{o}r\"{o}k, G., 2007, RevMexAA, 27, 8
\bibitem[]{} Belloni, T. M., Sanna, A., M\'{e}ndez, M., 2012, MNRAS, 426, 1701
\bibitem[]{} Bisnovatyi-Kogan, G. S., Ruzmaikin, A. A., 1974, Ap\&SS, 28, 45
\bibitem[]{} Bisnovatyi-Kogan, G. S., Ruzmaikin, A. A., 1976, Ap\&SS, 42, 401
\bibitem[]{} Blaes, O. M., Arras, P., Fragile, P. C., 2006, MNRAS, 369, 1235
\bibitem[]{} Chan, C.-K., 2009, ApJ, 704, 68
\bibitem[]{} Dolence, J. C., Gammie, C. F., Shiokawa, H., Noble, S. C., 2012, ApJ, 746, L10
\bibitem[]{} Done, C., Gierli\'{n}ski, M., Kubota, A., 2007, A\&AR, 15, 1
\bibitem[]{} Fu, W., Lai, D., 2009, ApJ, 690, 1386
\bibitem[]{} Fu, W., Lai, D., 2011, MNRAS, 410, 399
\bibitem[]{} Fu, W., Lai, D., 2012, MNRAS, 423, 831
\bibitem[]{} Fu, W., Lai, D., 2013, MNRAS, 431, 3697
\bibitem[]{} Henisey, K. B., Blaes, O. M., Fragile, P. C., Ferreira, B. T., 2009, ApJ, 706, 705
\bibitem[]{} Hor\'{a}k, J., Lai, D., 2013, MNRAS, 434, 2761
\bibitem[]{} Igumenshchev, I. V., Narayan, R., Abramowicz, M. A., 2003, ApJ, 592, 1042
\bibitem[]{} Kato, S., 1978, MNRAS, 185, 629
\bibitem[]{} Kato, S., Honma, F. \& Matsumoto, R. 1988a, MNRAS, 231, 37
\bibitem[]{} Kato, S., Honma, F. \& Matsumoto, R. 1988b, PASJ, 40, 709
\bibitem[]{} Kato, S., Wu, X.-B., Yang, L.-T. \& Yang, Z.-L. 1993, MNRAS, 260, 317   
\bibitem[]{} Kato, S., 2001, PASJ, 53, 1
\bibitem[]{} Lai, D., Tsang, D., 2009, MNRAS, 393, 979
\bibitem[]{} Lai, D., Fu, W., Tsang, D., Hor\'{a}k, J., Yu, C., 2013, IAUS, 290, 57
\bibitem[]{} Latter, H. N., Ogilvie, G. I., 2010, Icarus, 210, 318
\bibitem[]{} Lyubarskij, Y. E., Postnov, K. A., Prokhorov, M. E., 1994, MNRAS, 266, 583
\bibitem[]{} Mao, S. A., Psaltis, D., Milsom, J. A., 2009, ApJ, 703, 717
\bibitem[]{} McKinney, J. C., Tchekhovskoy, A., Blandford, R. D., 2012, MNRAS, 423, 3083
\bibitem[]{} Mignone, A., Bodo, G., Massaglia, S., Matsakos, T., Tesileanu, O., Zanni, C., Ferrari, A., 2007, ApJS, 170, 228
\bibitem[]{} Milsom, J. A., Taam, R. E., 1996, MNRAS, 283, 919
\bibitem[]{} Narayan, R., McClintock, J. E., 2012, MNRAS, 419, L69
\bibitem[]{} O'Neill, S. M., Reynolds, C. S., Miller, M. C., 2009, ApJ, 693, 1100
\bibitem[]{} O'Neill, S. M., Reynolds, C. S., Miller, M. C., Sorathia, K. A., 2011, ApJ, 736, 107
\bibitem[]{} Oda, H., Machida, M., Nakamura, K. E., Matsumoto, R., 2010, ApJ, 712, 639
\bibitem[]{} Ortega-Rodr\'{i}guez, M., Silbergleit, A., Wagoner, R., 2008, GApFD, 102, 75
\bibitem[]{} Ortega-Rodr\'{i}guez, M., Wagoner, R., 2000, ApJ, 537, 922 
\bibitem[]{} Paczy\'{n}ski, B., Witta, P. J., 1980, A\&A, 88, 23
\bibitem[]{} Papaloizou, J. C. B., Lin, D. N. C., 1988, ApJ, 331, 838
\bibitem[]{} Rein, H., Latter, H. N., 2013, MNRAS, 431, 145
\bibitem[]{} Remillard, R. A., McClintock, J. E., 2006, ARA\&A, 44, 49
\bibitem[]{} Reynolds, C. S., Miller, M. C., 2009, ApJ, 692, 869
\bibitem[]{} Rezzolla, L., Yoshida, S., Maccarone, T. J., Zanotti, O., 2003, MNRAS, 344, L37
\bibitem[]{} Rothstein, D. M., Lovelace, R. V. E., 2008, ApJ, 677, 1221
\bibitem[]{} Schmidt, J., Salo, H., Spahn, F., Petzschmann, O., 2001, Icarus, 153, 316
\bibitem[]{} Schmit, U., Tscharnuter, W. M., 1995, Icarus, 115, 304
\bibitem[]{} Schnittman, J. D., Bertschinger, E., 2004, ApJ, 606, 1098
\bibitem[]{} Shakura, N. I., Sunyaev, R. A., 1973, A\&A, 24, 337
\bibitem[]{} Shcherbakov, R. V., McKinney, J. C., 2013, ApJL, 774, L22
\bibitem[]{} Stella, L., Vietri, M., Morsink, S. M., 1999, ApJ, 524, L63
\bibitem[]{} Tagger, M., Varni\`{e}re, P., 2006, ApJ, 652, 1457
\bibitem[]{} Tsang, D., Lai, D., 2008, MNRAS, 387, 446
\bibitem[]{} Tsang, D., Lai, D., 2009a, MNRAS, 396, 589
\bibitem[]{} Tsang, D., Lai, D., 2009b, MNRAS, 400, 470
\bibitem[]{} de Val-Borro, M. et al., 2006, MNRAS, 370, 529
\bibitem[]{} Wellons, S., Zhu, Y., Psaltis, D., Narayan, R., McClintock, J. E., 2014, ApJ, 785, 142
\bibitem[]{} Yuan, F., Lin, J., Wu, K., Ho, L. C. 2009, MNRAS, 395, 2183
\end{thebibliography}
\end{document}